\documentclass[11pt]{article}

\usepackage[utf8]{inputenc}
\usepackage[none]{hyphenat} % añadido para evitar que LaTeX corte las palabras (poner \sloppy después de \begin{document})
\usepackage{amsmath}
\usepackage{amssymb}
\usepackage{amsfonts}
\usepackage{multirow}
\usepackage{enumerate}
\usepackage{hyperref}
\usepackage{graphicx}
\hypersetup{colorlinks=true, linkcolor=blue, urlcolor=blue, citecolor=red}
\usepackage{listings}
\usepackage{color}
\usepackage{rotating}
\usepackage[top=2cm, bottom=2cm, left=2cm, right=2cm]{geometry}
\usepackage{algorithm}
\usepackage{algorithmic}
\usepackage{longtable} % para tablas largas

\title{Stepwise regression revisited}
\author{Rom\'{a}n Salmer\'{o}n G\'{o}mez and Catalina Garc\'ia García}
\date{} %\date{\today}

\begin{document}

\renewcommand{\tablename}{Tabla}

  \maketitle

  \begin{abstract}
This paper shows that the degree of approximate multicollinearity in a linear regression model increases simply by including independent variables, even if these are not highly linearly related.
In the current situation where it is relatively easy to find linear models with a large number of independent variables, it is shown that this issue can lead to the erroneous conclusion that there is a worrying problem of approximate multicollinearity.
To avoid this situation, an adjusted variance inflation factor is proposed to compensate the presence of a large number of independent variables in the multiple linear regression model.
It is shown that this proposal has a direct impact on variable selection models based on influence relationships, which translates into a new decision criterion in the individual significance contrast to be considered in stepwise regression models or even directly in a multiple linear regression model.
  \end{abstract}


  \section{Introduction}

  Given the following multiple linear regression model:
  \begin{equation}
    \mathbf{y} = \mathbf{X}^{(k)} \boldsymbol{\beta}^{(k)} + \mathbf{u} = \beta_{1} + \beta_{2} \mathbf{X}_{2} + \cdots + \beta_{j} \mathbf{X}_{j} + \cdots + \beta_{k} \mathbf{X}_{k} + \mathbf{u},
    \label{modelo1}
  \end{equation}
for $n$ observations and $k$ independent variables, where $\mathbf{u}$ is the perturbation of the model that is assumed to be spherical with variance equal to $\sigma^{2}$, Appendix \ref{appendix0} shows that the degree of multicollinearity automatically increases when a new independent variable is added to the model. Specifically, it is shown that measures usually applied to detect this problem (such as the variance inflation factor, condition number or determinant of the correlation matrix) detect a greater degree of multicollinearity in the following model:
  \begin{eqnarray}
    \mathbf{y} &=& \mathbf{X}^{(k)} \boldsymbol{\beta}^{(k)} + \beta_{k+1} \mathbf{X}_{k+1} + \mathbf{u}  = \beta_{1} + \beta_{2} \mathbf{X}_{2} + \cdots + \beta_{k} \mathbf{X}_{k} + \beta_{k+1} \mathbf{X}_{k+1} + \mathbf{u} \nonumber \\
        &=& \mathbf{X}^{(k+1)} \boldsymbol{\beta}^{(k+1)} + \mathbf{u},
    \label{modelo2}
  \end{eqnarray}
than in model (\ref{modelo1}), regardless of whether the new variable included, $\mathbf{X}_{k+1}$, is strongly linearly related to those already existing in the model.

Focusing on the variance inflation factor (VIF), this situation can result in this measurement exceeding the threshold traditionally established as indicative of worrying multicollinearity, simply because there is a high number of independent variables in the linear model and not because the linear relationships are high.

At this point, two possibilities arise naturally: either modify the VIF or the threshold established as a concern so that the number of independent variables included in the model is taken into account.

Taking into account that the VIF is based on the calculation of a coefficient of determination of a linear model and that in econometrics there is already a factor that weights this measure penalizing the inclusion of variables such as the adjusted coefficient of determination  (see, for example, Gujarati \cite{Gujarati2003}, Johnston \cite{Johnston1984}, Novales \cite{Novales1993} or Wooldrigde \cite{Wooldrigde2013}), in the present work we opt for the first option, modifying the VIF considering a factor that penalizes the inclusion of variables so that the VIF increases as long as the linear relationships increase considerably. Thus, a new measurement is obtained which we will call the adjusted variance inflation factor (aVIF).

However, since the aVIF is considered merely from the point of view of detection, the enlarging effect of multicollinearity on the variance of the estimated coefficients of the model (see, for example, Curto and Pinto \cite{CurtoPinto2011}, Farrar and Glauber \cite{FarrarGlauber1967}, Gunst and Mason \cite{GunstMason1977}, Silvey \cite{Silvey1969} or Willian and Watts \cite{WillanWatts1978}) would persist and, by extension, so would its influence on the non-rejection of the null hypothesis in individual significance tests.

That is to say, to complete the work it is necessary to extend the correction of the aVIF to the individual significance tests. This will prevent situations in which a high VIF is obtained due to the number of independent variables included in the model (and not to the linear relationships existing between them) from concluding with a spurious non-rejection in the individual significance tests. Consequently, it is proposed to transfer in a reasoned way the correction made to the VIF with the aVIF to the experimental or theoretical value of the individual significance tests.

This proposal may be of interest in variable selection procedures of the \textit{stepwise regression} type (such as \textit{forward selection} or \textit{backward elimination}) based on the relevance of the variables to be included/eliminated and on their linear relationship with those already existing in the linear model. Also in situations where, after proposing an econometric model, it is decided to eliminate a variable simply because its associated VIF is higher than the threshold established as troubling instead of estiamting the model using alternative methodologies such as ridge regression (Hoerl and Kennard \cite{HoerlKennard1970a, HoerlKennard1970b}), LASSO (Tibshirani \cite{Tibshirani1996}), elastic-net (Zou and Hastie \cite{ZouHastie2005}) or raise regression (Salmerón, García and García \cite{SalmeronGarciaGarcia2024}).
Similarly, it could be applied on subset selection in which explanatory variables are deleted iteratively through the use of indicators for detecting multicollinearity, as the condition number or the variance inflation factor, see, for example, Tamura et al. \cite{Tamura2017} and Tamura et al. \cite{Tamura2019} and in the application of variable selection algorithms used in massive data sets (high $n$ and/or $k$ values) such as those proposed, for example, in Bingqing, Zhen, Jun and Cuiqing \cite{BingqingZhenJunCuiqing2022} or Lin, Foster and Ungar \cite{LinFosterUngar2001}.

The paper is organized as follows: section \ref{notation} summarizes the notation used in the present paper and highlights some interesting considerations about it. Section \ref{metohology}  defines the adjusted variance inflation factor (aVIF) from the variance inflation factor (VIF) using the corrected coefficient of determination in the auxiliary regression used to calculate the VIF. In addition, a Monte Carlo simulation shows that in a linear regression model in which the independent variables are generated independently, it is possible to use the VIF to indicate that the linear relationships are worrying. This situation would be avoided by using the aVIF. Section \ref{aVIF_properties} analyzes the main properties of the factor that adjusts the VIF resulting in the aVIF, among which it is noteworthy that the aVIF decreases with the inclusion of new variables in the linear model. Section \ref{connection} proposes an adjustment, based on the aVIF, in the decision rule of the individual significance tests that allows us to distinguish whether or not the non-rejection of the null hypothesis in this type of test is simply due to the fact that there is a high number of independent variables in the linear model. Finally, in section \ref{examples} the results discussed are illustrated by using simulated data to show how useful they are in step-by-step variable selection procedures, and in section \ref{conclusions} the main results obtained in this work are highlighted.

The code used in R \cite{RCoreTeam} to obtain the aforementioned results is available on Github at the web address \url{https://github.com/rnoremlas/aVIF}.

  \section{Notation and other interesting considerations}
    \label{notation}

In the present work the following matrix algebra notation will be used:
  \begin{itemize}
    \item $\mathbf{H}^{(p)}$ represents a matrix with $n$ rows and $p$ columns.
    \item $\mathbf{H}^{(p),t}$ represents the transposition of the matrix $\mathbf{H}^{(p)}$.
    \item $\mathbf{R}^{(p),-1}$ represents the inverse of the correlation matrix of $\mathbf{H}^{(p)}$.
    \item $\mathbf{H}^{(p)}_{-h}$ is the result of eliminating column $h$ from the matrix $\mathbf{H}^{(p)}$.
    \item $\mathbf{b}^{(p)}$ represents a column vector of $p$ elements.
    \item $\mathbf{b}^{(p),t}$ represents the transposition of the vector $\mathbf{b}^{(p)}$.
  \end{itemize}

On the other hand, see Marquardt and Snee \cite{MarquardtSnee1975}, Marquardt \cite{Marquardt1980}, Snee and Marquardt \cite{SneeMarquardt1984} or Salmerón, Rodríguez and García \cite{{Salmeron2020}} for more details, it is relevant to highlight the following two types of multicollinearity:
  \begin{itemize}
    \item Non-essential multicollinearity: linear relationship of the independent variables of the linear regression model with the constant term.
    \item Essential multicollinearity: linear relationship between the independent variables of the linear regression model excluding the constant term.
  \end{itemize}

It is important to make this distinction as the VIF (the measure on which this work is based) is only capable of detecting essential multicollinearity, completely ignoring non-essential multicollinearity (see, for example, Salmerón, García and García \cite{Salmeron2018}), so the use of this tool to detect the degree of existing multicollinearity is limited.

Another limitation to consider is that the VIF, being based on a coefficient of determination (see subsection \ref{defVIF}), is not suitable for calculating binary variables, in which case the use of non-linear models such as logit/probit is recommended. For this reason, all independent variables considered in this study are considered quantitative by default.

Finally, it should be noted that the adjusted variance inflation factor proposed is not related to the corrected variance inflation factor (cVIF) proposed in Curto and Pinto \cite{CurtoPinto2011}. In this case, the correction aims for a situation in which ``the real impact on variance can be overestimated by the traditional VIF when the explanatory variables contain no redundant information about the dependent variable''. Nor should it be confused with other VIF corrections whose objective is to correct the possible influence of the presence of outliers in the data (see, for example, Jacob and Varadharajan \cite{JacobVaradharajan2024}, Midi and Bagheri \cite{MidiBagheri2010} or Ekiz \cite{Ekiz2021}).

  \section{Methodology}
    \label{metohology}

  \subsection{Variance inflation factor}
    \label{defVIF}

The variance inflation factor (VIF) is one of the most commonly used measures to detect whether the degree of linear relationships (multicollinearity) in a linear regression model is troubling. This measure is calculated in association with each independent variable in the model (\ref{modelo1}), excluding the constant term, as:
  \begin{equation}
    VIF(j) = \frac{1}{1 - R_{j}^{2}}, \quad j=2,\dots,k,
    \label{VIF}
  \end{equation}
  where $R_{j}^{2}$ is the coefficient of determination of the following auxiliary regression:
  \begin{eqnarray}
    \mathbf{X}_{j} &=& \mathbf{X}_{-j}^{(k)} \boldsymbol{\alpha}^{(k)} + \mathbf{v} \nonumber \\
        &=& \alpha_{1} + \alpha_{2} \mathbf{X}_{2} + \dots + \alpha_{j-1} \mathbf{X}_{j-1} + \alpha_{j+1} \mathbf{X}_{j+1} + \dots + \alpha_{k} \mathbf{X}_{k} + \mathbf{v}, \label{reg_aux_1}
  \end{eqnarray}
  being $\mathbf{X}_{-j}^{(k)}$ the result of eliminating the variable $\mathbf{X}_{j}$ in $\mathbf{X}^{(k)}$.

That is to say, $R_{j}^{2}$ captures the linear relationship between $\mathbf{X}_{j}$ and the rest of the independent variables in the model (excluding the independent term), which is nothing more than the definition of approximate multicollinearity. The higher the value of $R_{j}^{2}$ (and, consequently, the higher the value of $VIF(j)$) the greater the existing linear relationship (multicollinearity).

Traditionally (see, for example, Marquardt \cite{Marquardt1970}), VIF values greater than 10 are considered to indicate that there is a worrying multicollinearity problem in the regression model.

  \subsubsection{Monte Carlo simulation}

As mentioned, Appendix \ref{appendix1} shows that the VIF of the model (\ref{modelo2}) is greater than that of the model (\ref{modelo1}). But how much greater will it be? This will depend on the linear relationship that exists between the variable that is added in the model (\ref{modelo2}) and the previously existing variables in the model (\ref{modelo1}).

In this section, a simulation is performed in which 40 independent variables are generated following a normal distribution where the mean can randomly take the values $\{ \pm 1, \pm 3, \pm 5 \}$ and the variance the values $\{ 1, 9, 15 \}$ (note that in this way non-essential multicollinearity is avoided). This calculation is made considering that the number of observations is equal to 50 and 150.

Next, in both cases, the variance inflation factor of the following 39 regressions is calculated:
  $$\mathbf{y} = \beta_{1} + \beta_{2} \mathbf{X}_{2} + \cdots + \beta_{j} \mathbf{X}_{j} + \mathbf{u}, \quad j=2,\dots,39.$$
Since the variables have been generated independently, it is to be expected that the degree of multicollinearity is not troubling.

\begin{figure}
    \centering
    \includegraphics[width=8.5cm]{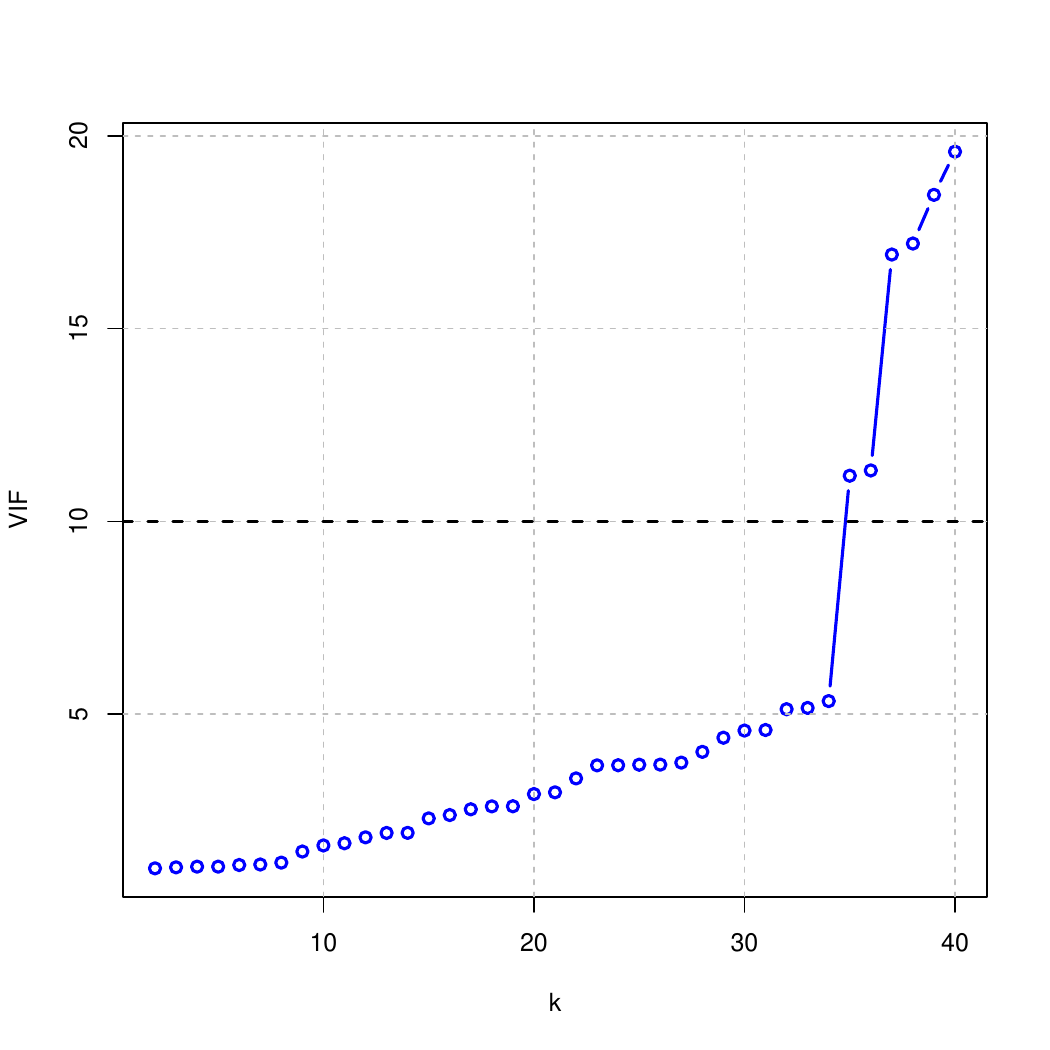}
    \includegraphics[width=8.5cm]{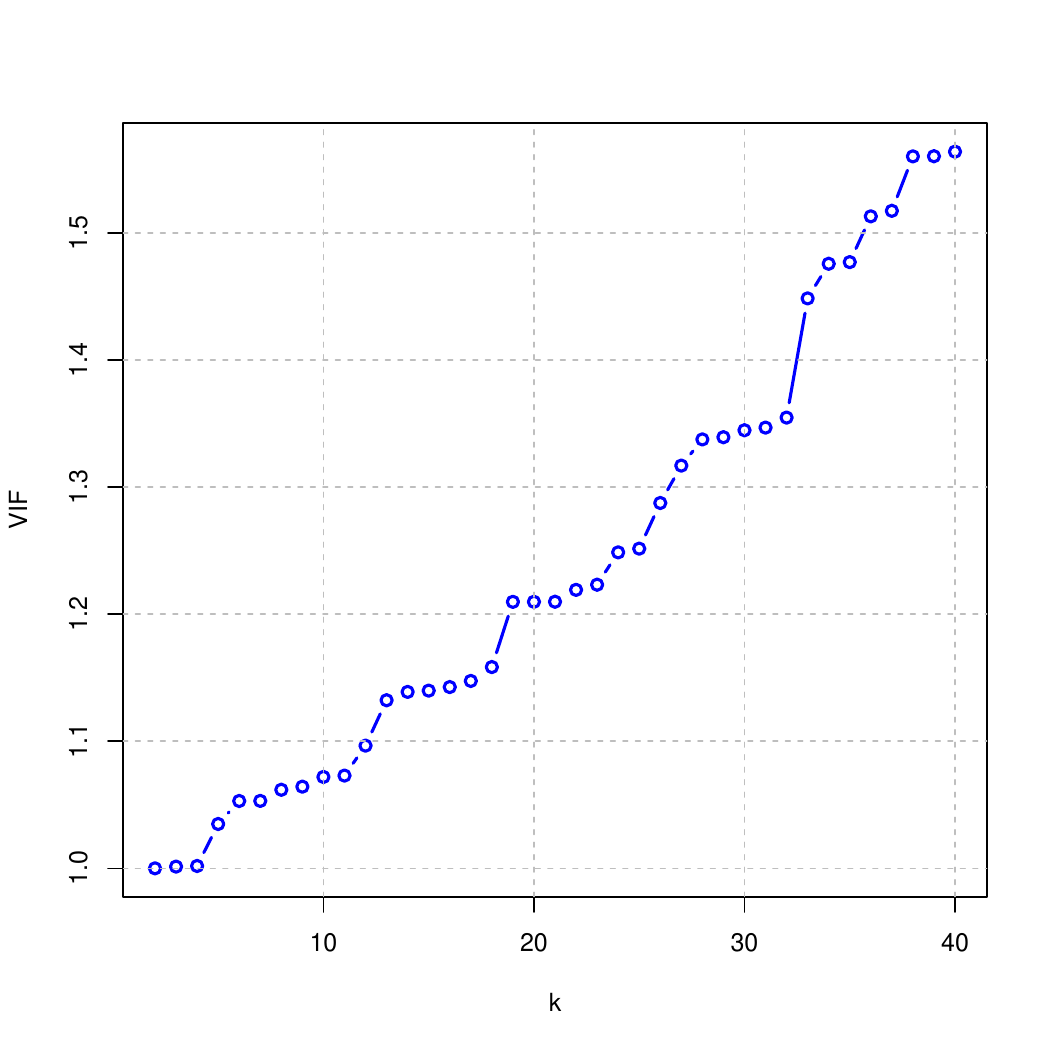}
    \caption{Maximum VIF as a function of the number of independent variables for 50 observations (left) and 150 observations (right)}\label{fig1}
  \end{figure}

However, as can be seen on the left side of Figure \ref{fig1}, for 50 observations the maximum VIF exceeds the threshold of 10 when the model has 35 independent variables (constant term included). That is to say, in a model with 50 observations and 35 independent variables generated completely independently, it would indicate that the degree of multicollinearity is worrying. Therefore, an erroneous conclusion would be reached simply due to the number of independent variables that the model contains and not due to the existing linear relationship between them.

However, when the number of observations is increased to 150, on the right side of Figure \ref{fig1}, it can be seen that the highest VIF does not exceed the value of 1.6. In this way, the role that the number of available observations plays in the problem of multicollinearity can be appreciated.

  \subsection{Adjusted variance inflation factor}

Among the tools used in the multiple linear regression model is the adjusted coefficient of determination. This measure has the same purpose as the coefficient of determination but taking into account (adjusting) the number of independent variables present in the model as well as the number of observations.

Therefore, the possibility of using this adjusted coefficient of determination to calculate the VIF from the expression (\ref{VIF}) arises naturally.

Thus, the adjusted variance inflation factor (aVIF) is defined, which is associated with each independent variable of the model (\ref{modelo1}), excluding the constant term, as:
  $$aVIF(j) = \frac{1}{1 - \overline{R}_{j}^{2}}, \quad j=2,\dots,k,$$
  where $\overline{R}_{j}^{2}$ is the adjusted coefficient of determination of the auxiliary regression (\ref{reg_aux_1}) and corresponds to the expression $\overline{R}_{j}^{2} = \left( 1 - \frac{n-1}{n-(k-1)} \right) \cdot (1 - R_{j}^{2})$.

  In fact, the aVIF can be expressed as:
  \begin{equation}
    aVIF(j) = \frac{n-k+1}{n-1} \cdot VIF(j) = a(n,k) \cdot VIF(j), \quad j=2,\dots,k,
    \label{aVIF}
  \end{equation}
  where $a(n,k) = \frac{n-k+1}{n-1}$ is a factor that weights the VIF according to the number of observations and independent variables present in the linear model (\ref{modelo1}).

  \subsubsection{Monte Carlo simulation}
        \label{simulation2}

\begin{figure}
    \centering
    \includegraphics[width=8.5cm]{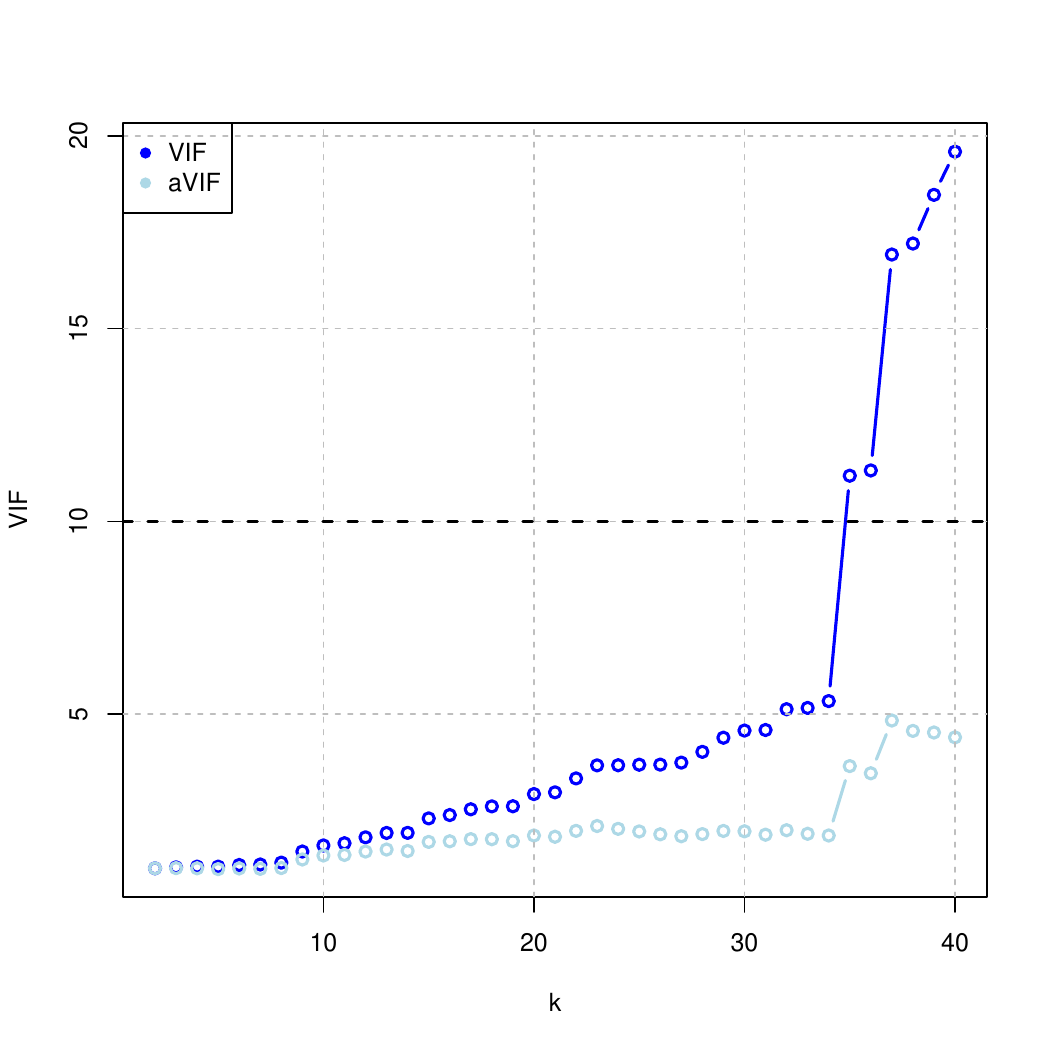}
    \includegraphics[width=8.5cm]{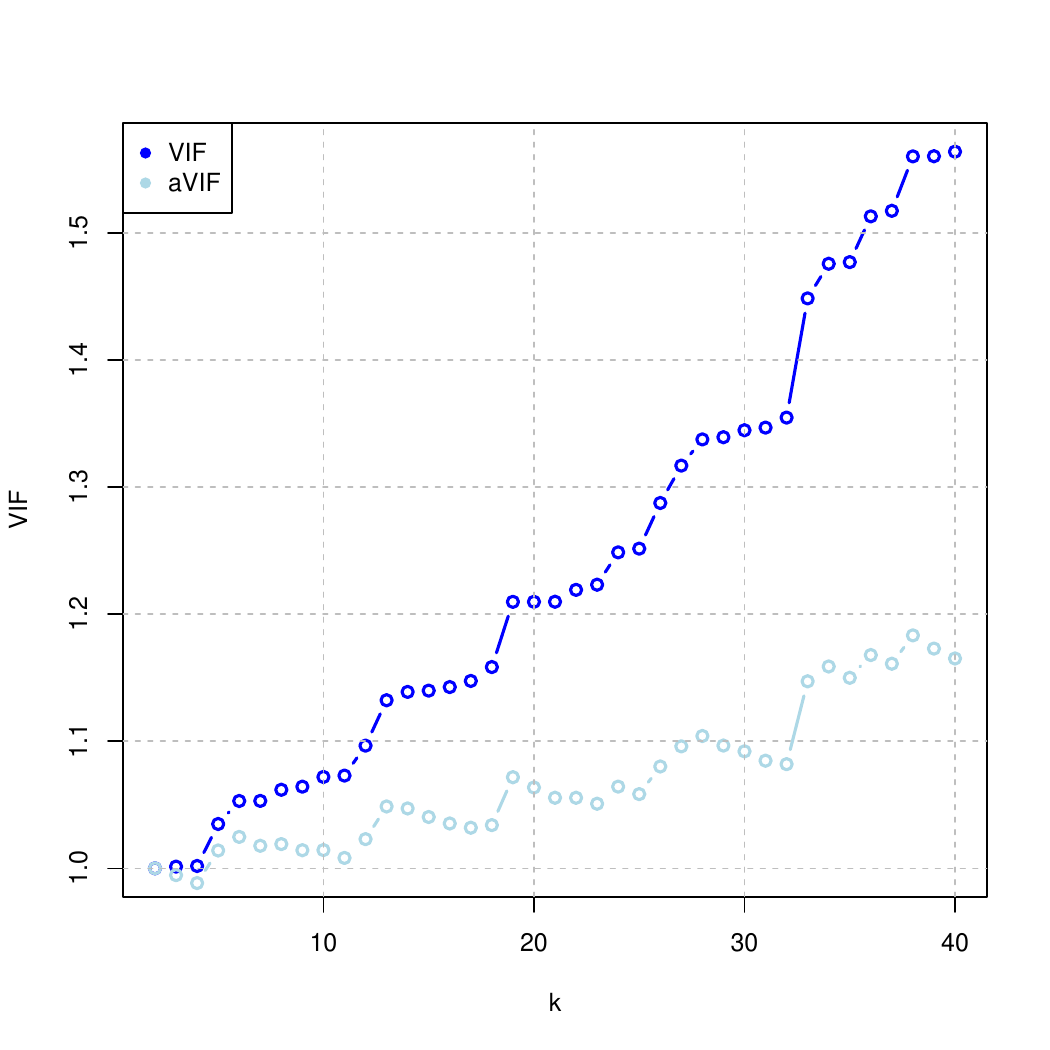}
    \caption{Maximum VIF and aVIF as a function of the number of independent variables for 50 observations (left) and 150 observations (right)}\label{fig2}
  \end{figure}

Repeating the simulation proposed in the previous subsection and calculating the aVIF according to the expression (\ref{aVIF}), we have the results shown in Figure \ref{fig2}. It can be seen that the reduction experienced in the aVIF implies (in the case where there are 50 observations) that all the values are below the threshold of 10 established as troubling.

  \section{aVIF properties}
    \label{aVIF_properties}

From the previous simulations, it can be seen that the aVIF mitigates the effect that the inclusion of independent variables has on the VIF. This question is formally studied below.

Considering the expression of the weighting factor $a(n,k)$, it can be seen that:
  \begin{enumerate}[a)]
    \item In the simple linear regression model ($k=2$), it is verified that $a(n,2) = 1$. Therefore, in this case $aVIF = VIF = 1$ (see Salmerón, Rodríguez and García \cite{Salmeron2020} for more details).
    \item If $k > 2$, then $n-k < n-2$ or, equivalently, $n-k+1 < n-2+1 = n-1$. Therefore, $a(n,k) = \frac{n-k+1}{n-1} < 1$ for any value of $n$ and $k>2$. Then $aVIF(j) < VIF(j)$ for any value of $n$ and $k>2$, with $j=2,\dots,k$. \\
        In addition, $VIF(j) - aVIF(j) = (1 - a(n,k)) \cdot VIF(j) = \frac{k-2}{n-1}  \cdot VIF(j)$ for $j=2,\dots,k$.
    \item If $n > k-1$ (which is always verified), then $a(n,k) > 0$.
    \item $\lim \limits_{n \rightarrow +\infty} aVIF(j) = VIF(j)$ due to $\lim \limits_{n \rightarrow +\infty} a(n,k) = 1$.
    \item When the independent variables are orthogonal (total absence of multicollinearity), it is verified that $VIF(j) = 1$ for all $j=2,\dots,k$. In that case, $aVIF(j) = \frac{n-k+1}{n-1} = a(n,k)$ for $j=2,\dots,k$. In short, the minimum value that the aVIF can take is less than one and always positive.  \\
        In addition, $VIF(j) - aVIF(j) = \frac{k-2}{n-1}$ for $j=2,\dots,k$, so that $VIF(j) - aVIF(j) > 10 \Leftrightarrow k > 2 + 10 \cdot (n-1)$, which is not feasible in the linear regression model estimated by Ordinary Least Squares (OLS) since in that case it is required that $n > k$.
    \item $a(n,k)$ increases when $n$ does for $k>2$ since then $\frac{\partial a(n,k)} {\partial n} = \frac{n-1 - n + k -1}{ (n-1)^{2}} = \frac{k-2}{ (n-1)^{2}} > 0$.
    \item $a(n,k)$ decreases as $k$ increases since $\frac{\partial a(n,k)}{\partial k} = - \frac{1}{n-1} < 0$.
  \end{enumerate}

In summary, $a(n,k)$ is the minimum value that the aVIF can take, which belongs to the interval $(0, 1)$ and increases when $n$ does and decreases when $k$ increases. Table \ref{tab1a} in Appendix \ref{appendix4} clearly illustrates these conclusions.

Therefore, the intended purpose is verified: the aVIF penalizes the inclusion of independent variables in the regression model in such a way that it will always have a lower value than the VIF.

Furthermore, since VIF is not affected by changes in origin or scale (see, for example, García, Salmerón, García and López-Martín \cite{Garcia2016}), neither is aVIF. Similarly, since VIF is only able to detect non-essential multicollinearity, aVIF has the same limitation.

  \subsection{Monte Carlo simulation}

A simulation is then performed in which data is generated for 250 variables $\mathbf{M}_{i} \sim N(\mu, \sigma)$ with $\mu, \sigma \in \{2, 3, 4, 5\}$ and for different sample sizes, specifically, it is considered that $n \in \{ 25, 50, 75, 100, 125, 150, 175, 200 \}$.

Then, for a fixed $\gamma \in \{ 0, 0.25, 0.5, 0.75, 0.8, 0.85, 0.9, 0.95 \}$, the following regressions are progressively proposed:
    \begin{eqnarray*}
        (\mbox{Model }1, \ k=3) \quad \mathbf{Y} &=& \beta_{1} + \beta_{2} \mathbf{X}_{2} + \beta_{3} \mathbf{X}_{3} + \mathbf{u}, \\
        (\mbox{Model }2, \ k=4) \quad \mathbf{Y} &=& \beta_{1} + \beta_{2} \mathbf{X}_{2} + \beta_{3} \mathbf{X}_{3} + \beta_{4} \mathbf{X}_{4} + \mathbf{u}, \\
        (\mbox{Model }3, \ k=5) \quad \mathbf{Y} &=& \beta_{1} + \beta_{2} \mathbf{X}_{2} + \beta_{3} \mathbf{X}_{3} + \beta_{4} \mathbf{X}_{4} + \beta_{5} \mathbf{X}_{5} + \mathbf{u}, \\
         & & \vdots \\
        (\mbox{Model }j, \ k=j) \quad\ \mathbf{Y} &=& \beta_{1} + \beta_{2} \mathbf{X}_{2} + \beta_{3} \mathbf{X}_{3} + \beta_{4} \mathbf{X}_{4} + \beta_{5} \mathbf{X}_{5} + \cdots + \beta_{j} \mathbf{X}_{j} + \mathbf{u} \nonumber \\
            &=& \mathbf{X}^{(j)} \boldsymbol{\beta}^{(j)} + \mathbf{u}, \\
         & & \vdots
    \end{eqnarray*}
    where the variables that form the matrix $\mathbf{X}^{(j)}$ are generated as follows:
    $$\mathbf{X}_{i} = \sqrt{1-\gamma^{2}} \cdot \mathbf{M}_{i} + \gamma \cdot \mathbf{M}_{j}, \ i=2,\dots, j.$$

This way of simulating data for the design matrix has been used previously in McDonald and Galarneau \cite{McDonaldGalarneau1975}, Wichern, D.W. and Churchill \cite{WichernChurchill1978}, Gibbons \cite{Gibbons1981}, Kibria \cite{Kibria2003} or Salmerón, García, López-Martín and García \cite{Salmeronetal2016} and aims to make the correlation between any two independent variables equal to $\gamma^{2}$.

For each model (1, 2, 3, etc.) the maximum VIF is calculated and the number of independent variables that makes it greater than 10 is obtained.
Table \ref{tab1} shows the results for all the values of $n$ and $\gamma$ considered.

It is observed that:
    \begin{itemize}
        \item The greater the value of $\gamma$, the lower the value of $k$.
        \item The greater the value of $n$, the greater the value of $k$.
    \end{itemize}
Both results were entirely predictable based on previous results.

\begin{table}
    \centering
    \begin{tabular}{ccccccccc}
        \hline
        $n$ vs $\gamma$ & 0 & 0.25 & 0.5 & 0.75 & 0.8 & 0.85 & 0.9 & 0.95 \\
        \hline
        25  &  19 &   19 &  13 &    4 &   4 &    3 &   3 &    3 \\
        50  &  39 &   39 &  30 &   18 &   9 &    5 &   4  &   3 \\
        75  &  61 &   59 &  36 &    7 &   7 &    4 &   3 &    3 \\
        100 &  78 &   79 &  68 &   19 &  10 &    4 &   4 &    3 \\
        125 & 105 &  106 &  98 &   70 &  49 &   12 &   7 &    4 \\
        150 & 117 &  116 &  90 &   13  &  8 &    8 &   4 &    3 \\
        175 & 142 &  143 & 132 &   74 &  28 &    6 &   5 &    3 \\
        200 & 165 &  161 & 146 &   40 &  11 &    6 &   3 &    3 \\
        \hline
    \end{tabular}
    \caption{Value of $k$ that makes the maximum VIF greater than 10} \label{tab1}
\end{table}

As previously mentioned, two options naturally arise that allow this behavior to be taken into account: either the traditionally used threshold of 10 is modified (increasing it) or the VIF is corrected (decreasing it).

Opting (obviously) for the second possibility, the previous simulation is repeated, in this case calculating the aVIF.
Table \ref{tab2} shows the values of $k$ that make the aVIF exceed the threshold of 10 for all the values of $n$ and $\gamma$ considered in the simulation. The values from Table \ref{tab1} are shown in brackets to facilitate comparison between the two measurements.

The observed behavior is the same as that found in the simulation in subsection \ref{simulation2}. In addition:
    \begin{itemize}
        \item The number of independent variables necessary to exceed the established threshold of 10 increases.
        \item When the linear relationship is low:
        \begin{itemize}
            \item There are cases when this increase means that for the fixed number of observations, the threshold of 10 is not exceeded, taking into account that it must be verified that $n>k$.
            \item In others, the value of $k$ is very close to $n$.
        \end{itemize}
        \item When $\gamma \geq 0.85$ the values of $k$ necessary to exceed the threshold practically coincide. That is to say, the use of the aVIF is of special interest (as was to be expected) when the linear relationships in the linear model are weak.
    \end{itemize}

\begin{table}
    \centering
    \begin{tabular}{ccccccccc}
        \hline
        $n$ vs $\gamma$ & 0 & 0.25 & 0.5 & 0.75 & 0.8 & 0.85 & 0.9 & 0.95 \\
        \hline
        25  &  NE (19) &   NE (19) &  22 (13) &  4  (4) & 4  (4) &  3  (3) & 3  (3) &  3  (3) \\
        50  &  NE (39) &  49 (39) & NE (30) &  20 (18) & 11  (9) &  5  (5) &  4 (4)  &  3 (3) \\
        75  &  NE (61) &  70 (59) & 61 (36) &   9 (7) &  7 (7) &  4  (4) &  3 (3) &   3 (3) \\
        100 &  99 (78) &  95 (79) & 95 (68) & 28  (19) & 10 (10) &  4  (4) &  4 (4) &  3  (3) \\
        125 & NE (105) &  NE (106) &  NE (98) &  NE  (70) & 119 (49) &  17 (12) &  7 (7) &  4  (4) \\
        150 & 147 (117) & NE  (116) & 136 (90) & 13  (13)  & 8 (8) &   8 (8) &  4 (4) &  3  (3) \\
        175 & NE (142) & 174 (143) & 174 (132) &  156 (74) & 93 (28) &  7  (6) & 5  (5) &  3  (3) \\
        200 & 195 (165) & 195 (161) & 193 (146) & 112  (40) & 12 (11) &  6  (6) &  3 (3) &  3  (3) \\
        \hline
    \end{tabular}
    \caption{Value of $k$ that makes the maximum aVIF (VIF in brackets) greater than 10 (NE means that there is no $k$ such that $k < n$ and $aVIF > 10$)} \label{tab2}
\end{table}

  \section{Connection with selection variable procedures}
    \label{connection}

As shown in the previous section, in models where there is a high number of independent variables, high VIF values can be obtained without the existing linear relationships (multicollinearity) being high. It has also been shown that this situation is corrected using the aVIF, making it possible to indicate that in situations where the VIF would suggest that the degree of multicollinearity is worrying, such a consideration would not be made if the aVIF were used.

However, since the individual significance tests of the linear model are affected by the VIF, even if the aVIF indicates the contrary, a high VIF value may mean that the null hypothesis is not rejected spuriously in this type of test.

In fact, given the linear regression model (\ref{modelo1}), the null hypothesis $H_{0}: \beta_{j}=0$ is rejected if:
  \begin{equation}
    t_{exp} = \left| \frac{\widehat{\beta}_{j}}{\sqrt{\widehat{var} (\widehat{\beta}_{j})}} \right| > t_{n-k}(1-\alpha/2), \quad j = 2,\dots,k,
    \label{t_exp}
  \end{equation}
  where $\widehat{\beta}_{j}$ is the OLS estimator of $\beta_{j}$, $\widehat{var} (\widehat{\beta}_{j}) = \frac{\widehat{\sigma}^{2}}{n \cdot var(\mathbf{X}_{j})} \cdot VIF(j)$ is the estimate of the variance of the coefficient estimators and $t_{n-k}(1-\alpha/2)$ is the value of a Student's t-distribution with $n-k$ degrees of freedom that leaves a probability of $1-\alpha/2$ on its left.

It is evident that a high VIF implies (with the exceptions mentioned by O'Brien \cite{OBrien2007}) a low $t_{exp}$ and, consequently, a tendency not to reject the null hypothesis of the coefficient being zero.

A possible solution to avoid this situation would be to use aVIF instead of VIF in the calculation of $t_{exp}$ which is equivalent to dividing $t_{exp}$ by $\sqrt{a(n,k)}$ or multiplying it by $b(n,k) = \frac{1}{\sqrt{a(n,k)}} = \sqrt{\frac{n-1}{n-k+1}}$:
  $$b(n,k) \cdot t_{exp} = \frac{t_{exp}}{\sqrt{a(n,k)}} = \left| \frac{\widehat{\beta}_{j}}{\sqrt{a(n,k) \cdot \widehat{var} (\widehat{\beta}_{j})}} \right| = \left| \frac{\widehat{\beta}_{j}}{\sqrt{\frac{\widehat{\sigma}^{2}}{n \cdot var(\mathbf{X}_{j})} \cdot aVIF(j)}} \right|.$$

By analogy with $a(n,k)$, it is evident that $b(n,k)$ increases when $k$ does (see Table \ref{tab2a} in Appendix \ref{appendix4}), so that $t_{exp}$ would be increased in such a way that the condition for rejecting the null hypothesis of the coefficient being zero could be verified.

Another possibility would be to multiply $t_{n-k}(1-\alpha/2)$ by $\sqrt{a(n,k)}$ since:
  \begin{equation}
    \frac{t_{exp}}{\sqrt{a(n,k)}} > t_{n-k}(1-\alpha/2) \Leftrightarrow t_{exp} > \sqrt{a(n,k)} \cdot t_{n-k}(1-\alpha/2),
    \label{at_exp}
  \end{equation}
where $\sqrt{a(n,k)}$ verifies that it decreases as $k$ increases (see Table \ref{tab3a} in Appendix \ref{appendix4}). That is to say, the increase of independent variables in the linear regression model would decrease the theoretical value with which to compare the experimental value in the individual significance tests.

Denoting $at_{n-k}(1-\alpha/2) = \sqrt{a(n,k)} \cdot t_{n-k}(1-\alpha/2)$, the following options can be considered:
  \begin{enumerate}[a)]
    \item As $\sqrt{a(n,k)} < \sqrt{1} = 1$ for $k>2$, then $at_{n-k}(1-\alpha/2) < t_{n-k}(1-\alpha/2)$, so if $t_{exp} > t_{n-k}(1-\alpha/2)$ initially, it is assured that $t_{exp} > at_{n-k}(1-\alpha/2)$. That is to say, if the null hypothesis is initially rejected, it will also be rejected with the adjustment, so it could be considered that the relationship between the independent variable in question and the dependent variable is so strong that it overcomes any effect that the degree of multicollinearity might have.
    \item In contrast, if the null hypothesis is not initially rejected, $t_{exp} < t_{n-k}(1-\alpha/2)$, as $at_{n-k}(1-\alpha/2) < t_{n-k}(1-\alpha/2)$, nothing can be said about the rejection or non-rejection of the null hypothesis once the adjustment has been made. In the case where the null hypothesis is not rejected in both cases, it could be considered that this is due to the lack of relationship between the independent variable in question and the dependent variable.
    \item Particularly interesting is the situation in which $t_{exp} < t_{n-k}(1-\alpha/2)$ initially, the null hypothesis of parameter nullity is not rejected, but $t_{exp} > at_{n-k}(1-\alpha/2)$ is verified, thus rejecting the null hypothesis after the adjustment made. In this case, the initial non-rejection of the null hypothesis could be due to the inflation of the variance of the estimators due to the high number of independent variables in the linear model and not so much to the degree of multicollinearity in the model.
  \end{enumerate}
This fact can be especially useful in variable selection procedures \textit{forward selection} or \textit{backward elimination} such as \textit{stepwise regression}, when the condition for including or eliminating a variable is based on criteria of individual significance of the coefficients of the independent variables instead of global criteria such as those used in model selection (for example, Akaike or Bayesian information criteria), where the inclusion of a greater number of independent variables is penalized in the sum of squares of the residuals of the linear model.

These types of models aim to develop models that are as interpretable as possible, thus avoiding the inclusion of independent variables with high linear relationships that can lead to unstable coefficient estimates (see, for example, Bertsimas and King \cite{BertsimasKing2016}). Using the decision rule given by the expression (\ref{at_exp}) instead of (\ref{t_exp}) can prevent variables with a high VIF from being excluded from the final model, due not so much to the existing multicollinearity but to the number of independent variables present in the model.

  \section{Example}
    \label{examples}

In this section, 50 observations are generated for 35 independent variables that have different levels of linear relationships with each other, in such a way as to analyze what would happen in the case of opting a) to eliminate variables whose coefficient is not significantly different from zero or b) to use \textit{stepwise} variable selection procedures in which the decision rules given by the expressions (\ref{t_exp}) or (\ref{at_exp}) are used in the individual significance tests. The above results are compared with those obtained when the sample is increased to 150 observations.

  \subsection{Monte Carlo simulation}
        \label{simulation_example}

Firstly, the data is generated. Specifically, $n=50$ observations are simulated for $k=35$ independent variables in such a way that the first is the constant term and the following 30 are generated independently according to:
  $$\mathbf{X}_{j} \sim N(\mu_{j}, \sigma^{2}_{j}), \quad j=2,\dots,31,$$
where $\mu_{j}$ and $\sigma_{j}$ take values at random from the sets $\{-10, -8, -6, -4, -2, 0, 2, 4, 6, 8, 10\}$ and $\{1, 2, 3, 4, 5\}$, respectively.

Once again, the values considered for $\sigma_{j}$ avoid the worrying presence of non-essential multicollinearity.

Table \ref{tab3} shows the values of $\mu_{j}$ and $\sigma_{j}$ applied in this simulation. With the aim of introducing a certain degree of essential multicolinearity into the linear model, the following four variables are generated from the previous variables:
  \begin{eqnarray*}
    \mathbf{X}_{32} &=& 4 \cdot \mathbf{X}_{2} - 3 \cdot \mathbf{X}_{3}  \cdot \mathbf{X}_{5} +  \mathbf{p}_{1}, \quad  \mathbf{p}_{1} \sim N(0, 2), \\
    \mathbf{X}_{33} &=& \mathbf{X}_{7} -  \mathbf{X}_{8} - \mathbf{p}_{2}, \quad  \mathbf{p}_{2} \sim N(0, 3), \\
    \mathbf{X}_{34} &=& 5 \cdot \mathbf{X}_{10} - 3 \cdot \mathbf{X}_{13} -  \mathbf{p}_{3}, \quad \mathbf{p}_{3} \sim N(0, 2), \\
    \mathbf{X}_{35} &=& \mathbf{X}_{15} +  \mathbf{X}_{17} + \mathbf{p}_{4}, \quad  \mathbf{p}_{4} \sim N(0, 3).
  \end{eqnarray*}
Thus, a linear relationship is expected between the variables $\mathbf{X}_{32}$ with $\mathbf{X}_{2}$, $\mathbf{X}_{3}$ and $\mathbf{X}_{5}$; $\mathbf{X}_{33}$ with $\mathbf{X}_{7}$ and $\mathbf{X}_{8}$; $\mathbf{X}_{34}$ with $\mathbf{X} _{10}$ and $\mathbf{X}_{13}$; and $\mathbf{X}_{35}$ with $\mathbf{X}_{15}$ and $\mathbf{X}_{17}$.

Table \ref{tab3} shows the VIFs associated with each variable (excepting the constant term), it can be seen that all the variables above have an associated VIF greater than 10 except $\mathbf{X}_{5}$, $\mathbf{X}_{7}$, $\mathbf{X}_{15}$, $\mathbf{X}_{17}$ and $\mathbf{X}_{35}$.

The above independent variables from the $\mathbf{X}^{(35)}$ matrix of the following linear model:
  \begin{equation}
    \mathbf{y} = \mathbf{X}^{(35)} \boldsymbol{\beta}^{(35)} + \mathbf{u}, \quad \mathbf{u} \sim N(0, 7),
    \label{modelo3}
  \end{equation}
where the coefficients $\boldsymbol{\beta}^{(35)}$ are randomly taken from the set $\{-7, -5, -3, -1, 0, 1, 3, 5, 7\}$.
Table \ref{tab3} shows the values of $\beta_{j}$ used in this simulation. Values equal to zero mean that the associated variable had no effect on the generation of the dependent variable.

In this case, it can be seen that the variables used in the generation of the dependent variable $\mathbf{y}$ are the constant term, $\mathbf{X}_{2}$, $\mathbf{X}_{3}$, $\mathbf{X}_{5}$, $\mathbf{X}_{8}$, $\mathbf{X}_{10}$, $\mathbf{X}_{12}$, $\mathbf{X}_{13}$, $\mathbf{X}_{14}$, $\mathbf{X}_{17}$, $\mathbf{X}_{18}$, $\mathbf{X}_{22}$, $\mathbf{X}_{25}$, $\mathbf{X}_{26}$, $\mathbf{X}_{27}$, $\mathbf{X}_{28}$, $\mathbf{X}_{30}$, $\mathbf{X}_{31}$, $\mathbf{X}_{32}$, $\mathbf{X}_{33}$ and $\mathbf{X}_{34}$.

This table also shows the decisions to be made according to the rule given in the expression (\ref{t_exp}). It can be seen that all the above variables have estimated coefficients significantly different from zero (at 5\% significance) except $\mathbf{X}_{2}$, $\mathbf{X}_{3}$, $\mathbf{X}_{8}$, $\mathbf{X}_{10}$, $\mathbf{X}_{13}$, $\mathbf{X}_{17}$ and $\mathbf{X}_{22}$.

Curiously, we can see that there is a variable, $\mathbf{X}_{4}$, with a coefficient set at 0 and which, therefore, has not been used to generate the dependent variable and, nevertheless, has an associated estimated coefficient significantly different from zero. The rest of the variables with coefficients set at zero do not have estimated coefficients significantly different from zero.

\begin{sidewaystable}
    \centering
    {\small
    \begin{tabular}{ccccccccccccc}
        \hline
        $j$ & $\mu_{j}$ & $\sigma_{j}$ & $\beta_{j}$ & VIF & aVIF & $VIF - aVIF$ & $t_{exp}$ & $t_{n-k}(1-\alpha/2)$ & $t_{exp} > t_{n-k}(1-\alpha/2)$ & $at_{n-k}(1-\alpha/2)$ & $t_{exp} > at_{n-k}(1-\alpha/2)$ & Option \\
        \hline
        1 &  &  & 3 &  &  &  & 0.005 & 2.131 & No & 1.218 & No & b) \\
        \textbf{2} & \textbf{-4} & \textbf{2} & \textbf{1} & \textbf{64.415} & \textbf{22.348} & \textbf{42.067} & \textbf{1.361} & \textbf{2.131} & \textbf{No} & \textbf{1.218} & \textbf{Yes} & \textbf{c)} \\
        3 & -8 & 2 & 1 & 48.744 & 16.911 & 31.833 & 0.667 & 2.131 & No & 1.218 & No & b) \\
        4 & -2 & 2 & 0 & 1.961 & 0.68 & 1.281 & 2.69 & 2.131 & Yes & 1.218 & Yes & a) \\
        5 & -8 & 1 & 5 & 7.595 & 2.635 & 4.96 & 3.623 & 2.131 & Yes & 1.218 & Yes & a) \\
        6 & -6 & 5 & 0 & 3.299 & 1.144 & 2.154 & 0.011 & 2.131 & No & 1.218 & No & b) \\
        7 & -6 & 5 & 0 & 8.957 & 3.108 & 5.85 & 0.13 & 2.131 & No & 1.218 & No & b) \\
        \textbf{8} & \textbf{6} & \textbf{3} & \textbf{-1} & \textbf{10.276} & \textbf{3.565} & \textbf{6.711} & \textbf{1.351} & \textbf{2.131} & \textbf{No} & \textbf{1.218} & \textbf{Yes} & \textbf{c)} \\
        9 & 4 & 1 & 0 & 3.267 & 1.133 & 2.133 & 0.364 & 2.131 & No & 1.218 & No & b) \\
        10 & 10 & 5 & 3 & 277.997 & 96.448 & 181.549 & 0.02 & 2.131 & No & 1.218 & No & b) \\
        11 & 4 & 2 & 0 & 3.275 & 1.136 & 2.139 & 0.32 & 2.131 & No & 1.218 & No & b) \\
        12 & 0 & 5 & -3 & 2.603 & 0.903 & 1.7 & 11.63 & 2.131 & Yes & 1.218 & Yes & a) \\
        \textbf{13} & \textbf{8} & \textbf{4} & \textbf{1} & \textbf{81.937} & \textbf{28.427} & \textbf{53.51} & \textbf{1.915} & \textbf{2.131} & \textbf{No} & \textbf{1.218} & \textbf{Yes} & \textbf{c)} \\
        14 & 6 & 2 & 5 & 2.749 & 0.954 & 1.795 & 7.409 & 2.131 & Yes & 1.218 & Yes & a) \\
        15 & -8 & 2 & 0 & 3.729 & 1.294 & 2.435 & 0.289 & 2.131 & No & 1.218 & No & b) \\
        16 & -4 & 5 & 0 & 2.59 & 0.898 & 1.691 & 0.727 & 2.131 & No & 1.218 & No & b) \\
        \textbf{17} & \textbf{-4} & \textbf{1} & \textbf{5} & \textbf{4.281} & \textbf{1.485} & \textbf{2.796} & \textbf{1.36} & \textbf{2.131} & \textbf{No} & \textbf{1.218} & \textbf{Yes} & \textbf{c)} \\
        18 & 8 & 5 & 5 & 2.756 & 0.956 & 1.8 & 24.955 & 2.131 & Yes & 1.218 & Yes & a) \\
        19 & 2 & 1 & 0 & 3.441 & 1.194 & 2.247 & 0.469 & 2.131 & No & 1.218 & No & b) \\
        20 & -6 & 5 & 0 & 3.667 & 1.272 & 2.395 & 0.385 & 2.131 & No & 1.218 & No & b) \\
        21 & 0 & 4 & 0 & 3.946 & 1.369 & 2.577 & 0.011 & 2.131 & No & 1.218 & No & b) \\
        22 & 2 & 1 & -1 & 5.034 & 1.746 & 3.287 & 0.877 & 2.131 & No & 1.218 & No & b) \\
        23 & 2 & 3 & 0 & 3.111 & 1.079 & 2.031 & 0.907 & 2.131 & No & 1.218 & No & b) \\
        2\textbf{4} & \textbf{6} & \textbf{4} & \textbf{0} & \textbf{4.167} & \textbf{1.446} & \textbf{2.721} & \textbf{1.256} & \textbf{2.131} & \textbf{No} & \textbf{1.218} & \textbf{Yes} & \textbf{c)} \\
        25 & -8 & 1 & -5 & 3.227 & 1.12 & 2.108 & 2.757 & 2.131 & Yes & 1.218 & Yes & a) \\
        26 & 2 & 4 & -3 & 3.122 & 1.083 & 2.039 & 8.809 & 2.131 & Yes & 1.218 & Yes & a) \\
        27 & -6 & 5 & 3 & 2.746 & 0.953 & 1.793 & 11.49 & 2.131 & Yes & 1.218 & Yes & a) \\
        28 & -10 & 4 & -7 & 2.379 & 0.825 & 1.554 & 24.505 & 2.131 & Yes & 1.218 & Yes & a) \\
        29 & 6 & 3 & 0 & 3.526 & 1.223 & 2.303 & 0.575 & 2.131 & No & 1.218 & No & b) \\
        30 & -8 & 5 & 7 & 4.065 & 1.41 & 2.655 & 25.498 & 2.131 & Yes & 1.218 & Yes & a) \\
        31 & 4 & 4 & 1 & 3.572 & 1.239 & 2.333 & 2.955 & 2.131 & Yes & 1.218 & Yes & a) \\
        32 &  &  & -7 & 122.265 & 42.418 & 79.846 & 10.431 & 2.131 & Yes & 1.218 & Yes & a) \\
        33 &  &  & 5 & 14.513 & 5.035 & 9.478 & 12.589 & 2.131 & Yes & 1.218 & Yes & a) \\
        34 &  &  & 1 & 339.852 & 117.908 & 221.944 & 2.977 & 2.131 & Yes & 1.218 & Yes & a) \\
        35 &  &  & 0 & 4.347 & 1.508 & 2.839 & 0.802 & 2.131 & No & 1.218 & No & b) \\
        \hline
    \end{tabular}
    }
    \caption{Values of $\mu_{j}$, $\sigma_{j}$, $\beta_{j}$, $VIF(j)$, $aVIF(j)$, $VIF(j)-aVIF(j)$, $t_{exp}$, $t_{n-k}(1-\alpha/2)$ and $at_{n-k}(1-\alpha/2)$ for each simulated variable} \label{tab3}
\end{sidewaystable}

  \subsection{Elimination of variables}

In the introduction it is mentioned that one of the solutions used when proposing regression models to avoid the problem of multicollinearity is the elimination of variables. To avoid eliminating variables with VIFs inflated by the number of independent variables, as mentioned above, the situation of interest in this case would be the one classified as option c). In Table \ref{tab3} it can be seen that there are 5 independent variables ($\mathbf{X}_{2}$, $\mathbf{X}_{8}$, $\mathbf{X}_{13}$, $\mathbf{X} _{17}$ and $\mathbf{X}_{24}$, highlighted in bold) in this situation, two of them having VIF values well above 10 (together with a significant reduction in the same reflected in the aVIF) and the first four having been used in the generation of $\mathbf{y}$.

Therefore, five variables have been detected that in these cases would be discarded for not having an estimated coefficient significantly different from zero and that, nevertheless, once the number of independent variables existing in the model has been taken into account through the factor $a(50.35) = \frac{16}{49} = 0{.}326$, they would have it.

Thus, Table \ref{tab4} shows the OLS estimation of both the model (\ref{modelo3}) and the final model once the variables with a coefficient not significantly different from zero have been eliminated according to the expression (\ref{at_exp}) instead of (\ref{t_exp}). It can be seen that the variables $\mathbf{X}_{2}$, $\mathbf{X}_{8}$, $\mathbf{X}_{13}$, $\mathbf{X}_{17}$ and $\mathbf{X}_{24}$ have coefficients that are significantly different from zero. Furthermore, it can be seen that the model in which 15 variables have been eliminated has a higher corrected coefficient of determination (and a lower value in Akaike's model selection criterion) than the model in which all variables are taken into account, and a lower value in Akaike's model selection criterion.

  \subsection{Stepwise regression}

In Bertsimas and King \cite{BertsimasKing2016}, as an alternative to the direct elimination of independent variables, an iterative process is proposed in which variables with a coefficient not significantly different from zero are eliminated, starting with the one with the lowest experimental value, $t_{exp}$. The process ends when all the variables remaining in the model have a coefficient significantly different from zero.

Establishing this procedure using the rule given in (\ref{t_exp}) to determine which coefficients are significantly different from zero leads to the same model previously referred to as the \textit{elimination} model in Table \ref{tab4}.

Whereas if the rule given in (\ref{at_exp}) is used, the variables $\mathbf{X}_{22}$ and $\mathbf{X}_{23}$ are added to the previous model (\textit{elimination} model).
These variables do not verify the rule given in (\ref{t_exp}) or in (\ref{at_exp}) when all the variables are considered (see Table \ref{tab3}), however they do verify the rule given in (\ref{at_exp}), although not the one given in (\ref{t_exp}), when the commented step-by-step procedure is established.

The results of the estimation of this \textit{stepwise} model are also shown in Table \ref{tab4}. It can be seen that the coefficients of the variables $\mathbf{X}_{22}$ and $\mathbf{X}_{23}$ are significantly different from zero at a significance level of 10\%.

There may be some controversy over the suitability of relaxing the level of significance from 5\% to 10\%; however, it can be seen that this model offers the highest corrected coefficient of determination and the lowest value in the Akaike information criterion (AIC) of the three estimates. That is to say, the model obtained by the step-by-step procedure is the most preferable of all, indicating that it is worth including the variables $\mathbf{X}_{22}$ and $\mathbf{X}_{23}$.

Finally, to ensure that the coefficient of determination is between zero and one, in the three models in Table \ref{tab4} the constant term is included even if it is not significantly different from zero.

\begin{longtable}{ccccc} % https://minisconlatex.blogspot.com/2012/01/como-hacer-tablas-largas-que-ocupen.html
                         % https://es.overleaf.com/latex/examples/a-longtable-example/xxwzfxkxxjmc
  %\centering
  %\begin{tabular}{ccc} %lD{.}{.}{3}D{.}{.}{3}}
    \hline
     & $\beta_{j}$ & Initial model & Elimination model & Stepwise model \\ %\multicolumn{1}{c}{Model 1} &  \multicolumn{1}{c}{Model 2}\\
    \hline
    \endfirsthead
    \hline
     & $\beta_{j}$ & Initial model & Final model & Stepwise model \\ %\multicolumn{1}{c}{Model 1} &  \multicolumn{1}{c}{Model 2}\\
    \hline
    \endhead
    \hline
    \multicolumn{5}{c}{Continued on the next page.}
    \endfoot
    \endlastfoot
    Intercept & 3 & 0.128 & -4.768 & 7.394 \\
                 &  & (24.272) & (11.601) & (13.049) \\
    $X_{2}$ & 1       & 4.084 & 1.550$^{*}$& 2.060**  \\
                 &  & (3.001) & (0.602) & (0.618) \\
    $X_{3}$ & 1       & -1.724 & &  \\
                 &  & (2.585) & & \\
    $X_{4}$ &  0      & 1.595$^{*}$ & 1.680$^{***}$&  1.661*** \\
                 &  & (0.593) & (0.446)& (0.421) \\
    $X_{5}$ &  5      & 7.112$^{**}$ & 5.490$^{***}$& 6.463*** \\
                 &  & (1.963) & (0.804)& (0.840)\\
    $X_{6}$ & 0       & 0.004 & &  \\
                 &  & (0.314) & & \\
    $X_{7}$ & 0       & 0.057 & &  \\
                 &  & (0.435) & & \\
    $X_{8}$ & -1       & -1.036 & -1.232$^{***}$& -1.204*** \\
                 &  & (0.767) & (0.300)& (0.278) \\
    $X_{9}$ & 0       & 0.528 & &  \\
                 &  & (1.452) & & \\
    $X_{10}$ & 3       & -0.055 & &  \\
                 &  & (2.758) & & \\
    $X_{11}$ & 0      & 0.189 & &  \\
                 &  & (0.589) & & \\
    $X_{12}$ & -3      & -2.857$^{***}$ & -2.832$^{***}$& -2.931*** \\
                 &  & (0.246) & (0.162)& (0.160) \\
    $X_{13}$ & 1      & 2.942 & 2.881$^{***}$& 2.843*** \\
                 &  & (1.536) & (0.216)& (0.200) \\
    $X_{14}$ & 5      & 5.224$^{***}$ & 5.698$^{***}$& 5.360*** \\
                 &  & (0.705) & (0.440)& (0.429) \\
    $X_{15}$ & 0      & -0.196 & &  \\
                 &  & (0.678) & & \\
    $X_{16}$ & 0      & 0.181 & &  \\
                 &  & (0.248) & & \\
    $X_{17}$ & 5      & 2.172 & 3.833$^{***}$& 3.304*** \\
                 &  & (1.597) & (0.945)& (0.895) \\
    $X_{18}$ & 5      & 5.621$^{***}$ & 5.478$^{***}$& 5.517***  \\
                 &  & (0.225) & (0.137)& (0.127) \\
    $X_{19}$ &  0     & 0.660 & &  \\
                 &  & (1.408) & & \\
    $X_{20}$ & 0      & 0.118 & &  \\
                 &  & (0.306) & & \\
    $X_{21}$ & 0      & -0.004 & &  \\
                 &  & (0.372) & & \\
    $X_{22}$  & -1     & -1.333 & & -1.666$^{\bullet}$ \\
                 &  & (1.520) & & (0.864) \\
    $X_{23}$ &  0     & 0.362 & & 0.466$^{\bullet}$ \\
                 &  & (0.400) & & (0.234) \\
    $X_{24}$ & 0      & -0.520 & -0.744$^{**}$&  -0.606* \\
                 &  & (0.414) & (0.222)& (0.244) \\
    $X_{25}$  &  -5    & -4.040$^{*}$ & -4.267$^{***}$& -4.423*** \\
                 &  & (1.466) & (0.862)& (0.838) \\
    $X_{26}$ &  -3     & -2.845$^{***}$ & -2.931$^{***}$& -2.912*** \\
                 &  & (0.323) & (0.210)& (0.194) \\
    $X_{27}$ &  3     & 2.748$^{***}$ & 2.684$^{***}$& 2.802*** \\
                 &  & (0.239) & (0.148)& (0.150) \\
    $X_{28}$ &  -7     & -6.851$^{***}$ & -6.964$^{***}$& -6.852*** \\
                 &  & (0.280) & (0.190)& (0.182) \\
    $X_{29}$ &  0     & 0.295 & &  \\
                 &  & (0.513) & & \\
    $X_{30}$  &  7    & 6.599$^{***}$ & 6.511$^{***}$& 6.644*** \\
                 &  & (0.259) & (0.158)& (0.176) \\
    $X_{31}$ &  1     & 1.185$^{**}$ & 1.073$^{***}$& 1.221*** \\
                 &  & (0.401) & (0.235)& (0.226) \\
    $X_{32}$ &  -7     & -7.566$^{***}$ & -7.057$^{***}$& -7.134*** \\
                 &  & (0.725) & (0.122)& (0.117) \\
    $X_{33}$ &  5     & 5.050$^{***}$ & 4.979$^{***}$& 5.014*** \\
                 &  & (0.401) & (0.128)& (0.119) \\
    $X_{34}$ &  1     & 1.609$^{**}$ & 1.600$^{***}$& 1.605*** \\
                 &  & (0.540) & (0.037)& (0.035) \\
    $X_{35}$ &  0     & 0.290 & \\
                 &  & (0.361) & & \\
    \hline
    $n$ & &  50     &   50  &  50 \\
    $k$ & &  35     &   20   &  22 \\
    $\widehat{\sigma}$ & &   5.338 &    4.632& 4.27 \\
    $R^{2}$ & &   0.9993 &    0.9989 & 0.9992 \\
    $\overline{R}^{2}$ & &  0.9977  &  0.9983 & 0.9986 \\
    AIC & &  321.187  & 311.649  & 304.207 \\
    $F_{exp}$ & & 639.855 & 1520.383 & 1614.371 \\
    $p-value$ of $F_{exp}$ & &   0.000 &    0.000 & 0.000 \\
    \hline
%    \multicolumn{3}{c}{Significance: %{p{.7\linewidth}}{Significance:
%                  $*** \equiv p < 0{.}001$;
%                  $** \equiv p < 0{.}01$;
%                  $* \equiv p < 0{.}05$}\\
%    \hline
  %\end{tabular}
  \caption{OLS estimation of the model simulated in the example section (significance: $*** \equiv p < 0{.}001$; $** \equiv p < 0{.}01$; $* \equiv p < 0{.}05$; $\bullet \equiv p < 0{.}1$)} \label{tab4}
\end{longtable}

  \subsection{Effect of sample size}

In section \ref{aVIF_properties} it can be seen that the increase in $n$ compensates for the reduction in $a(n,k)$ and $\sqrt{a(n,k)}$ when $k$ increases. Thus, if the simulation proposed in subsection \ref{simulation_example} is repeated considering that $n=150$, only one variable ($\mathbf{X}_{25}$) would be in the situation referred to as option c).

This is surely due to the fact that $a(n,k) \rightarrow 1$ when $n \rightarrow +\infty$ or $n >>> k$ and to the effect that the number of observations has on the degree of linear relationships:
  \begin{itemize}
    \item In the model with 50 observations, the maximum VIF and aVIF are equal to 339'852 and 117'908, respectively.
    \item whereas in the model with 150 observations they are equal to 58'987 and 46'319.
  \end{itemize}
  In both cases, these would be the values associated with the $\mathbf{X}_{34}$ variable.

This ``healing'' effect of a large sample size on the problem of multicollinearity can also be seen in the work of Salmerón, García and García \cite{Salmeron2018}, where, among other questions, it is shown by using different simulations that the greater the number of observations, the fewer the number of condition.

  \section{Conclusions}
    \label{conclusions}

With the aim of obtaining models that are as interpretable as possible, the criteria for the step-by-step variable selection procedures include selecting independent variables that are as orthogonal as possible to each other. Thus, for example, Tabachnick and Fidell \cite{TabachnickFidell2001} recommend that independent variables with a simple linear correlation greater than 0.7 should not be included in multiple linear regression analysis.

Other works, such as that of Bertsimas and King \cite{BertsimasKing2016}, highlight the limitation of using simple correlation as a criterion and warn of the convenience of using tools that allow the measurement of linear relationships of more than two variables, proposing the use of the condition number.

This work focuses on the effect that having a high number of independent variables can have on the degree of multicollinearity of a linear regression model, showing that it can be established that this problem is a concern because there are many independent variables and not so much because of the linear relationships that exist between them.

Specifically, based on the variance inflation factor (VIF) and the corrected coefficient of determination, a coefficient is proposed that corrects this multicolinearity detection tool by adjusting its value according to the number of observations and independent variables of the linear model.  Since the VIF forms part of the experimental value used in the contrasts of individual significance of the coefficients of the independent variables of the model, it is immediate to apply this adjustment to the decision rule that allows us to determine if the estimation of these coefficients can be considered significantly different from zero (or not) at a given level of significance, simply by multiplying the theoretical value of the corresponding Student's t-distribution by the values of $\sqrt{a(n, k)} = \sqrt{\frac{n-k-1}{n-1}}$ determined by the observations, $n$, and number of independent variables, $k$, available in the specific model being analyzed.

Finally, it is shown that using this adjustment in stepwise regression procedures can help to obtain preferable models (with a higher corrected coefficient of determination or a lower value in the Akaike information criterion) than if the traditional decision rule is used (especially) when the number of independent variables is high and not too far from the number of available observations.

  %\printbibliography
  %\bibliographystyle{Chicago} % We choose the "plain" reference style
  %\bibliography{bib} % Entries are in the refs.bib file

\begin{thebibliography}{99}
	\bibitem{Belsley1991} Belsley, D.A. (1991) Conditioning diagnostics: collinearity and weak data in regression. NewYork: John Wiley.
	\bibitem{Belsley2005} Belsley, D.A., Kuh E. and Welsch, R.E. (2005) Regression diagnostics: identifying influential data and sources of collinearity. New York (NY): John Wiley and Sons.
    \bibitem{BertsimasKing2016} Bertsimas, D. and King, A. (2016). OR Forum — An Algorithmic Approach to Linear Regression. Operations Research, 64 (1), 2--16.
    \bibitem{BingqingZhenJunCuiqing2022} Bingqing, L., Zhen, P., Jun, Z. and Cuiqing, C. (2022). Fast feature selection via streamwise procedure for massive data. Brazilian Journal of Probability and Statistics, 36 (1), 81--102.
    \bibitem{CurtoPinto2011} Curto,  J.D. and Pinto, J.C. (2011). The corrected vif (cvif). Journal of Applied Statistics, 38 (7), 1499--1507.
	\bibitem{DelHierro2021} Del Hierro, A., García, C.B. and Salmerón, R. (2021). Analysis of the condition number in the raise regression. Communications in Statistics - Theory and Methods, 50(24), 6195--6210.
    \bibitem{Ekiz2021} Ekiz, O.U. (2021). An improved robust variance inflation factor: Reducing the negative effects of good leverage points. Kuwait Journal of Science, 50 (2A), 1--5.
    \bibitem{FarrarGlauber1967} Farrar, D.E. and Glauber, R.R. (1967). Multicollinearity in regression analysis: the problem revisited. The Review of Economic and Statistics, 49 (1), 92--107.
    \bibitem{Garcia2016} García, J., Salmerón, R., García, C.B. and López-Martín, M.M. (2016). Standardization of Variables and Collinearity Diagnostic in Ridge Regression. International Statistical Review, 84(2), 245--266.
    \bibitem{Garcia2019} García, C., Salmerón, R. and García, C.B. (2019). Choice of the ridge factor from the correlation matrix determinant. Journal of Statistical Computation and Simulation, 89(2), 211--231.
    \bibitem{Gibbons1981} Gibbons, D.G. (1981). A simulation study of some ridge estimators. Journal of American Statistical Association, 76, pp. 131--139.
    \bibitem{Gujarati2003} Gujarati, D.N. (2003). Basic Econometrics. McGraw-Hill (fourth edition).
    \bibitem{GunstMason1977} Gunst, R.L. and Mason, R.L. (1977). Advantages of examining multicollinearities in regression analysis. Biometrics, 33 (1), 249--260.
    \bibitem{HoerlKennard1970a} Hoerl, A.E. and Kennard, R.W. (1970a). Ridge Regression: Biased Estimation for Nonorthogonal Problems. Technometrics, 12 (1), 55--67.
    \bibitem{HoerlKennard1970b} Hoerl, A.E. and Kennard, R.W. (1970b). Ridge Regression: Applications to Nonorthogonal Problems. Technometrics, 12 (1), 69--82.
    \bibitem{JacobVaradharajan2024} Jacob, J. and Varadharajan, R. (2024). Robust Variance Inflation Factor: A Promising Approach for Collinearity Diagnostics in the Presence of Outliers. Sankhya B 86, 845--871.
    \bibitem{Johnston1984} Johnston, J. (1972). Econometric Methods. McGraw-Hill.
    \bibitem{Kibria2003} Kibria, B. (2003). Performance of some new ridge regression estimators. Communications in Statistics - Simulation and Computation, 32 (2), pp. 419--435.
    \bibitem{LinFosterUngar2001} Lin, D., Foster, D.P. and Ungar, L.H. (2011). VIF regression: a fast regression algorithm for large data. Journal of the American Statistical Association, 106 (493), 232--247.
    \bibitem{Marquardt1970} Marquardt, D. W. (1970). Generalized inverse, ridge regression, biased linear estimation and nonlinear estimation.  Technometrics, 12 (3), 591--612.
    \bibitem{Marquardt1980} Marquardt, D.W. (1980). You should standardize the predictor variables in your regression models. Journal of the American Statistical Association, 75 (369), 87--91.
    \bibitem{MarquardtSnee1975} Marquardt, D.W. and Snee, R. (1975). Ridge regression in practice. The American Statistician, 29 (1), 3--20.
    \bibitem{McDonaldGalarneau1975} McDonald, G.C. and Galarneau, D.I. (1975). A Monte Carlo evaluation of some ridge type estimators. Journal of American Statistical Association, 70, pp. 407--416.
    \bibitem{MidiBagheri2010} Midi, H. and Bagheri, A. (2010). Robust multicollinearity diagnostic measure in collinear data set. Proceedings of the 4th international conference on applied mathematics, simulation, modeling, 138--142.
    \bibitem{Novales1993} Novales, A. (1993). Econometría. McGraw-Hill.
    \bibitem{OBrien2007} O'Brien, R. M. (2007). A caution regarding rules of thumb for variance inflation factors. Quality and quantity, 41 (5), 673--690.
    \bibitem{RCoreTeam} R Core Team (2022) R: A Language and Environment for Statistical Computing, R Foundation for Statistical Computing, Vienna, Austria.
    \bibitem{Salmeronetal2016} Salmerón, R., García, J., López-Martín, M.M. and García, C.B. (2016). Collinearity diagnostic in ridge estimation through the variance inflation factor. Journal of Applied Statistics, 43 (10), pp. 1831--1849.
    \bibitem{Salmeron2018} Salmerón, R., García, C.B. and García, J. (2018). Variance Inflation Factor and Condition Number in multiple linear regression. Journal of Statistical Computation and Simulation, 88(12), 2365--2384.
    \bibitem{Salmeron2020} Salmerón, R., Rodríguez, A. and García, C.B. (2020). Diagnosis and quantification of the non-essential collinearity. Computational Statistics, 35(2), 647--666.
    \bibitem{SalmeronGarciaGarcia2024} Salmerón, R., García, C.B. and García, J. (2024). The raise regression: justification, properties and application. International Statistical Review (online), doi: https://doi.org/10.1111/insr.12575.
    \bibitem{Silvey1969} Silvey, S. (1969). Multicollinearity and imprecise estimation. Journal of the Royal Statistical Society. Series B (Methodological), 31 (3), 539--552.
    \bibitem{SneeMarquardt1984} Snee, R.D. and Marquardt D.W. (1984). Collinearity diagnostics depend on the domain of prediction, the model and the data. The American Statistician, 38 (2), 83--87.
    \bibitem{TabachnickFidell2001} Tabachnick, B.G. and Fidell, L.S. (2001). Using Multivariate Statistics 4th ed. Allyn and Bacon, Boston.
    \bibitem{Tamura2017} Tamura, R., Kobayashi, K., Takano, Y., Miyashiro, R., Nakata, K. and Matsui, T. (2017). Best subset selection for eliminating multicollinearity. Journal of the Operations Research Society of Japan, 60(3), 321-336.
    \bibitem{Tamura2019} Tamura, R., Kobayashi, K., Takano, Y., Miyashiro, R., Nakata, K. and Matsui, T. (2019). Mixed integer quadratic optimization formulations for eliminating multicollinearity based on variance inflation factor. Journal of Global Optimization, 73, 431-446.
    \bibitem{Tibshirani1996} Tibshirani, R. (1996). Regression shrinkage and selection via the LASSO. Journal of the Royal Statistical Society. Series B (Methodological), 58 (1), 267--288.
    \bibitem{WichernChurchill1978} Wichern, D.W. and Churchill, G.A. (1978). A comparison of ridge estimators. Technometrics, 20, pp. 301--311.
    \bibitem{WillanWatts1978} Willan, A.R. and Watts, D.G. (1978). Meaningful multicollinearity measures. Technometrics, 20 (4), 407--412.
    \bibitem{Wooldrigde2013} Wooldrigde, J.M. (2013). Introductory Econometrics. A Modern Approach. SOUTH-WESTERN, CENGAGE Learning (fifth edition).
    \bibitem{ZouHastie2005} Zou, H. and Hastie, T. (2005). Regularization and variable selection via the elastic net. Journal of the Royal Statistical Society: Series B (Statistical Methodology), 67 (2), 301--320.
\end{thebibliography}

  \appendix

  \section{Multicollinearity and the inclusion of variables}
    \label{appendix0}

  \subsection{From the variance inflation factor}
    \label{appendix1}

In order to establish that the degree of multicollinearity in the model (\ref{modelo2}) is greater than that of the model (\ref{modelo1}) based on the variance inflation factor, it is necessary to demonstrate that the coefficient of determination, denoted as $R_{k+1,j}^{2}$, of the following auxiliary regression:
    \begin{eqnarray}
        \mathbf{X}_{j} &=& \alpha_{1} + \alpha_{2} \mathbf{X}_{2} + \dots + \alpha_{j-1} \mathbf{X}_{j-1} + \alpha_{j+1} \mathbf{X}_{j+1} + \dots + \alpha_{k} \mathbf{X}_{k}  + \alpha_{k+1} \mathbf{X}_{k+1} + \mathbf{v}
        \label{reg_aux_2} \\
        &=& \mathbf{X}_{-j}^{(k+1)} \boldsymbol{\alpha}^{(k+1)} + \mathbf{v}, \nonumber
    \end{eqnarray}
is greater than that of the auxiliary regression (\ref{reg_aux_1}), denoted from now on as $R_{k,j}^{2}$.

Although it is well known that by including independent variables in the linear regression model the coefficient of determination increases even if the included variables are not relevant (which would directly imply that $R_{k+1,j}^{2} > R_{k,j}^{2}$), it is worth demonstrating this point.

Since models (\ref{reg_aux_1}) and (\ref{reg_aux_2}) have the same dependent variable, the total sum of squares will be the same in both models, so proving that $R_{k+1,j}^{2} > R_{k,j}^{2}$ is equivalent to proving that $SCR_{k} > SCR_{k+1}$, where $SCR_{k}$ is the sum of squares of the residuals of model (\ref{reg_aux_1}) and $SCR_{k+1}$ that of model (\ref{reg_aux_2}).

In fact, taking into account that  $\mathbf{X}_{-j}^{(k+1)} = [\mathbf{X}_{-j}^{(k)} \ \ \mathbf{X}_{k+1}]$, it is verified that the estimator of model (\ref{reg_aux_2}) is:
    \begin{eqnarray}
        \widehat{\boldsymbol{\alpha}}^{(k+1)} &=& \left( \mathbf{X}_{-j}^{(k+1), t} \mathbf{X}_{-j}^{(k+1)} \right)^{-1} \mathbf{X}_{-j}^{(k+1),t} \mathbf{X}_{j} = \left(
                \begin{array}{cc}
                    \mathbf{X}_{-j}^{(k),t} \mathbf{X}_{-j}^{(k)} & \mathbf{X}_{-j}^{(k),t} \mathbf{X}_{k+1} \\
                    \mathbf{X}_{k+1}^{t} \mathbf{X}_{-j}^{(k)} & \mathbf{X}_{k+1}^{t} \mathbf{X}_{k+1}
                \end{array} \right) \cdot \left(
                \begin{array}{c}
                    \mathbf{X}_{-j}^{(k),t} \mathbf{X}_{j} \\
                    \mathbf{X}_{k+1}^{t} \mathbf{X}_{j}
                \end{array} \right) \nonumber \\
            &=& \left(
                \begin{array}{cc}
                    \mathbf{A} & \mathbf{B} \\
                    \mathbf{B}^{t} & \mathbf{C}
                \end{array} \right) \cdot \left(
                \begin{array}{c}
                    \mathbf{X}_{-j}^{(k),t} \mathbf{X}_{j} \\
                    \mathbf{X}_{k+1}^{t} \mathbf{X}_{j}
                \end{array} \right). \nonumber
    \end{eqnarray}

Consequently,
    \begin{eqnarray}
        \widehat{\boldsymbol{\alpha}}^{(k+1),t} \mathbf{X}_{-j}^{(k+1),t} \mathbf{X}_{j} &=& \left( \mathbf{X}_{j}^{t} \mathbf{X}_{-j}^{(k)} \ \ \ \mathbf{X}_{j}^{t} \mathbf{X}_{k+1}  \right) \cdot \left(
                \begin{array}{cc}
                    \mathbf{A} & \mathbf{B} \\
                    \mathbf{B}^{t} & \mathbf{C}
                \end{array} \right) \cdot \left(
                \begin{array}{c}
                    \mathbf{X}_{-j}^{(k),t} \mathbf{X}_{j} \\
                    \mathbf{X}_{k+1}^{t} \mathbf{X}_{j}
                \end{array} \right) \nonumber \\
        &=& \left( \mathbf{X}_{j}^{t} \mathbf{X}_{-j}^{(k)} \cdot \mathbf{A} + \mathbf{X}_{j}^{t} \mathbf{X}_{k+1} \cdot \mathbf{B}^{t} \ \ \ \mathbf{X}_{j}^{t} \mathbf{X}_{-j}^{(k)} \cdot \mathbf{B} + \mathbf{X}_{j}^{t} \mathbf{X}_{k+1} \cdot \mathbf{C} \right) \cdot \left(
                \begin{array}{c}
                    \mathbf{X}_{-j}^{(k),t} \mathbf{X}_{j} \\
                    \mathbf{X}_{k+1}^{t} \mathbf{X}_{j}
                \end{array} \right) \nonumber \\
        &=& \mathbf{X}_{j}^{t} \mathbf{X}_{-j}^{(k)} \cdot \mathbf{A} \cdot \mathbf{X}_{-j}^{(k),t} \mathbf{X}_{j} + \mathbf{X}_{j}^{t} \mathbf{X}_{k+1} \cdot \mathbf{B}^{t} \cdot \mathbf{X}_{-j}^{(k),t} \mathbf{X}_{j} \nonumber \\
        & & + \mathbf{X}_{j}^{t} \mathbf{X}_{-j}^{(k)} \cdot \mathbf{B} \cdot \mathbf{X}_{k+1}^{t} \mathbf{X}_{j} + \mathbf{X}_{j}^{t} \mathbf{X}_{k+1} \cdot \mathbf{C} \cdot \mathbf{X}_{k+1}^{t} \mathbf{X}_{j}. \label{expr1}
    \end{eqnarray}

Due to:
    \begin{eqnarray}
        \mathbf{C} &=& \left( \mathbf{X}_{k+1}^{t} \mathbf{X}_{k+1} - \mathbf{X}_{k+1}^{t} \left( \mathbf{X}_{-j}^{(k),t} \mathbf{X}_{-j}^{(k)} \right)^{-1} \mathbf{X}_{-j}^{(k),t} \mathbf{X}_{k+1} \right)^{-1} = \left( SCR_{k}^{\delta} \right)^{-1}, \nonumber \\
        \mathbf{B} &=& - \left( \mathbf{X}_{-j}^{(k),t} \mathbf{X}_{-j}^{(k)} \right)^{-1} \mathbf{X}_{-j}^{(k),t} \mathbf{X}_{k+1} \cdot \mathbf{C} = - \widehat{\boldsymbol{\delta}}^{(k)} \cdot \left( SCR_{k}^{\delta} \right)^{-1}, \nonumber \\
        \mathbf{A} &=& \left( \mathbf{X}_{-j}^{(k),t} \mathbf{X}_{-j}^{(k)} \right)^{-1} + \left( \mathbf{X}_{-j}^{(k),t} \mathbf{X}_{-j}^{(k)} \right)^{-1} \mathbf{X}_{-j}^{(k),t} \mathbf{X}_{k+1} \cdot \mathbf{C} \cdot \mathbf{X}_{k+1} \mathbf{X}_{-j}^{(k),t} \left( \mathbf{X}_{-j}^{(k),t} \mathbf{X}_{-j}^{(k)} \right)^{-1} \nonumber \\
            &=& \left( \mathbf{X}_{-j}^{(k),t} \mathbf{X}_{-j}^{(k)} \right)^{-1} + \frac{\widehat{\boldsymbol{\delta}}^{(k)} \widehat{\boldsymbol{\delta}}^{(k),t}}{SCR_{k}^{\delta}}, \nonumber
    \end{eqnarray}
   where $\widehat{\boldsymbol{\delta}}^{(k)}$ and $SCR_{k}^{\delta}$ are, respectively, the estimator of the coefficients and the sum of squares of the residuals of the following regression:
    \begin{eqnarray}
        \mathbf{X}_{k+1} &=& \mathbf{X}_{-j}^{(k)} \boldsymbol{\delta}^{(k)} + \mathbf{w}
        \label{reg_aux_3} \\
        &=& \delta_{1} + \delta_{2} \mathbf{X}_{2} + \dots + \delta_{j-1} \mathbf{X}_{j-1} + \delta_{j+1} \mathbf{X}_{j+1} + \dots + \delta_{k} \mathbf{X}_{k} + \mathbf{w}, \nonumber
    \end{eqnarray}
   the expression (\ref{expr1}) can be rewritten as:
    \begin{eqnarray}
        \widehat{\boldsymbol{\alpha}}^{(k+1),t} \mathbf{X}_{-j}^{(k+1),t} \mathbf{X}_{j} &=&  \mathbf{X}_{j}^{t} \mathbf{X}_{-j}^{(k)} \cdot \left( \mathbf{X}_{-j}^{(k),t} \mathbf{X}_{-j}^{(k)} \right)^{-1} \cdot 			\mathbf{X}_{-j}^{(k),t} \mathbf{X}_{j} +
			\frac{1}{SCR_{k}^{\delta}} \mathbf{X}_{j}^{t} \mathbf{X}_{-j}^{(k)} \cdot \widehat{\boldsymbol{\delta}}^{(k)} \widehat{\boldsymbol{\delta}}^{(k),t} \cdot \mathbf{X}_{-j}^{(k),t} \mathbf{X}_{j} \nonumber \\
		& & - \frac{1}{SCR_{k}^{\delta}}  \mathbf{X}_{j}^{t} \mathbf{X}_{k+1} \cdot \widehat{\boldsymbol{\delta}}^{(k),t} \cdot \mathbf{X}_{-j}^{(k),t} \mathbf{X}_{j}
			- \frac{1}{SCR_{k}^{\delta}}  \mathbf{X}_{j}^{t} \mathbf{X}_{-j}^{(k)} \cdot \widehat{\boldsymbol{\delta}}^{(k)} \cdot \mathbf{X}_{k+1}^{t} \mathbf{X}_{j} \nonumber \\
	        & & + \frac{1}{SCR_{k}^{\delta}} \mathbf{X}_{j}^{t} \mathbf{X}_{k+1} \cdot \mathbf{X}_{k+1}^{t} \mathbf{X}_{j} \nonumber \\
		&=& \widehat{\boldsymbol{\alpha}}^{(k),t} \cdot \mathbf{X}_{-j}^{(k),t} \mathbf{X}_{j}
			+ \frac{1}{SCR_{k}^{\delta}} \mathbf{X}_{j}^{t} \left( \mathbf{X}_{-j}^{(k)} \cdot \widehat{\boldsymbol{\delta}}^{(k)} \widehat{\boldsymbol{\delta}}^{(k),t} \cdot \mathbf{X}_{-j}^{(k),t} - \mathbf{X}_{k+1} 					\cdot \widehat{\boldsymbol{\delta}}^{(k),t} \cdot \mathbf{X}_{-j}^{(k),t} \right. \nonumber \\
		& & \left. - \mathbf{X}_{-j}^{(k)} \cdot \widehat{\boldsymbol{\delta}}^{(k)} \cdot \mathbf{X}_{k+1}^{t} + \mathbf{X}_{k+1} \cdot \mathbf{X}_{k+1}^{t} \right) \mathbf{X}_{j} \nonumber \\
		&=& \widehat{\boldsymbol{\alpha}}^{(k),t} \cdot \mathbf{X}_{-j}^{(k),t} \mathbf{X}_{j}
			+ \frac{1}{SCR_{k}^{\delta}} \mathbf{X}_{j}^{t} \left(  \mathbf{X}_{k+1} - \mathbf{X}_{-j}^{(k)} \widehat{\boldsymbol{\delta}}^{(k)} \right) \left(  \mathbf{X}_{k+1} - \mathbf{X}_{-j}^{(k)} 									\widehat{\boldsymbol{\delta}}^{(k)} \right)^{t} \mathbf{X}_{j}. \nonumber
    \end{eqnarray}

Considering that $\mathbf{e}_{k}^{\delta} = \mathbf{X}_{k+1} - \mathbf{X}_{-j}^{(k)} \widehat{\boldsymbol{\delta}}^{(k)}$ are the residuals of the model (\ref{reg_aux_3}), the previous expression is equivalent to:
$$\widehat{\boldsymbol{\alpha}}^{(k+1),t} \mathbf{X}_{-j}^{(k+1),t} \mathbf{X}_{j} = \widehat{\boldsymbol{\alpha}}^{(k),t} \cdot \mathbf{X}_{-j}^{(k),t} \mathbf{X}_{j} + \frac{(\mathbf{X}_{j}^{t} \mathbf{e}_{k}^{\delta})^{2}}{SCR_{k}^{\delta}},$$
and, consequently, it follows that:
    \begin{eqnarray}
        SCR_{k+1} &=& \mathbf{X}_{j}^{t} \mathbf{X}_{j} - \widehat{\boldsymbol{\alpha}}^{(k+1),t} \mathbf{X}_{-j}^{(k+1),t} \mathbf{X}_{j} =
		\mathbf{X}_{j}^{t} \mathbf{X}_{j} - \widehat{\boldsymbol{\alpha}}^{(k),t} \cdot \mathbf{X}_{-j}^{(k),t} \mathbf{X}_{j} - \frac{(\mathbf{X}_{j}^{t} \mathbf{e}_{k}^{\delta})^{2}}{SCR_{k}^{\delta}} \nonumber \\
			&=& SCR_{k}  - \frac{(\mathbf{X}_{j}^{t} \mathbf{e}_{k}^{\delta})^{2}}{SCR_{k}^{\delta}}. \nonumber
    \end{eqnarray}

	Since $\frac{(\mathbf{X}_{j}^{t} \mathbf{e}_{k}^{\delta})^{2}}{SCR_{k}^{\delta}} > 0$, then it is proven that $SCR_{k+1} < SCR_{k}$.

  \subsection{From the condition number}
    \label{appendix2}

Given the model (\ref{modelo1}), following Belsley \cite{Belsley1991} and Belsley, Kuh and Welsch \cite{Belsley2005}, the condition number is given by:
	$$CN \left( \mathbf{X}^{(k)} \right) = \sqrt{\frac{\lambda_{max}^{(k)}}{\lambda_{min}^{(k)}}},$$
	where $\lambda_{max}^{(k)}$ and $\lambda_{min}^{(k)}$ are, respectively, the maximum and minimum eigenvalues of the matrix $\mathbf{X}^{(k)}$ once transformed, if it is not, into unit length.

On the other hand, the condition number of the model (\ref{modelo2}) is given by:
	$$CN(\mathbf{X}^{(k+1)}) = \sqrt{\frac{\lambda_{max}^{(k+1)}}{\lambda_{min}^{(k+1)}}},$$
	where $\lambda_{max}^{(k+1)}$ and $\lambda_{min}^{(k+1)}$ are, respectively, the maximum and minimum eigenvalues of the matrix $\mathbf{X}^{(k+1)}$ once transformed, if it is not, into unit length.

Taking into account that $\mathbf{X}^{(k+1)} = [ \mathbf{X}^{(k)} \ \ \mathbf{X}_{k+1} ]$, it is verified that:
	$$\mathbf{X}^{(k+1),t} \mathbf{X}^{(k+1)} = \left(
		\begin{array}{cc}
			\mathbf{X}^{(k),t} \mathbf{X}^{(k)} & \mathbf{X}^{(k),t} \mathbf{X}_{k+1} \\
			\mathbf{X}_{k+1}^{t} \mathbf{X}^{(k)} & \mathbf{X}_{k+1}^{t} \mathbf{X}_{k+1}
		\end{array} \right).$$

In that case, as $\mathbf{X}^{(k,t)} \mathbf{X}^{(k)}$ is a submatrix of $\mathbf{X}^{(k+1,t)} \mathbf{X}^{(k+1)}$ it holds (see Theorem A by Del Hierro, García and Salmerón \cite{DelHierro2021}, page 6208) that:
	$$0 < \lambda_{min}^{(k+1)} \leq \lambda_{min}^{(k)} \leq \cdots \leq \lambda_{max}^{(k)} \leq \lambda_{max}^{(k+1)},$$
	and then:
	$$\frac{1}{\lambda_{min}^{(k)}} < \frac{1}{\lambda_{min}^{(k+1)}}, \quad
    \frac{\lambda_{min}^{(k+1)}}{\lambda_{min}^{(k)}} < 1 < \frac{\lambda_{max}^{(k)}}{\lambda_{min}^{(k)}} < \frac{\lambda_{max}^{(k+1)}}{\lambda_{min}^{(k)}}.$$
    %, \quad 1 < \frac{\lambda_{max}^{(k+1)}}{\lambda_{max}^{(k)}}.$$

From the first chain of inequalities, it can be seen that:
    $$\frac{\lambda_{max}^{(k+1)}}{\lambda_{min}^{(k)}} < \frac{\lambda_{max}^{(k+1)}}{\lambda_{min}^{(k+1)}},$$
and, consequently, taking the second into account:
	$$\frac{\lambda_{max}^{(k)}}{\lambda_{min}^{(k)}} < \frac{\lambda_{max}^{(k+1)}}{\lambda_{min}^{(k)}} < \frac{\lambda_{max}^{(k+1)}}{\lambda_{min}^{(k+1)}},$$
that is, $CN \left( \mathbf{X}^{(k)} \right) < CN(\mathbf{X}^{(k+1)})$. This implies that the degree of multicollinearity in model (\ref{modelo2}) is greater than in model (\ref{modelo1}).

  \subsection{From the determinant of the correlation matrix}
    \label{appendix3}

Considering that $\mathbf{R}^(k)$ and $\mathbf{R}^(k+1)$ are, respectively, the correlation matrices associated with $\mathbf{X}^{(k)}$ and $\mathbf{X}^{(k+1)} = [ \mathbf{X}^{(k)} \ \ \mathbf{X}_{k+1} ]$, it is verified that:
	$$\mathbf{R}^{(k+1)} = \left(
		\begin{array}{cc}
			\mathbf{R}^{(k)} & \mathbf{R}_{\mathbf{X}^{(k)}, \mathbf{X}_{k+1}} \\
			\mathbf{R}_{\mathbf{X}^{(k)}, \mathbf{X}_{k+1}} & \mathbf{R}_{\mathbf{X}_{k+1}, \mathbf{X}_{k+1}}
		\end{array} \right),$$
	where $\mathbf{R}_{\mathbf{X}^{(k)}, \mathbf{X}_{k+1}}$ presents the correlations of $\mathbf{X}^{(k)}$ with $\mathbf{X}_{k+1}$ and $\mathbf{R}_{\mathbf{X}_{k+1}, \mathbf{X}_{k+1}} = 1$.

In this case, the determinant of the matrix $\mathbf{R}^{(k+1)}$ is equal to:
	$$| \mathbf{R}^{(k+1)} | = | \mathbf{R}^{(k)} | \cdot | 1 - \mathbf{R}_{\mathbf{X}^{(k)}, \mathbf{X}_{k+1}}^{t} \mathbf{R}^{(k),-1} \mathbf{R}_{\mathbf{X}^{(k)}, \mathbf{X}_{k+1}} | =  | \mathbf{R}^{(k)} | \cdot \left( 1 - \mathbf{R}_{\mathbf{X}^{(k)}, \mathbf{X}_{k+1}}^{t} \mathbf{R}^{(k),-1} \mathbf{R}_{\mathbf{X}^{(k)}, \mathbf{X}_{k+1}} \right),$$
where it has been used that $\mathbf{R}^t_{\mathbf{X}^{(k)}, \mathbf{X}_{k+1}} \mathbf{R}^{(k),-1} \mathbf{R}_{\mathbf{X}{(k)}, \mathbf{X}_{k+1}}$ is a scalar.

Suppose that $1 - \mathbf{R}_{\mathbf{X}^{(k)}, \mathbf{X}_{k+1}}^{t} \mathbf{R}^{(k),-1} \mathbf{R}_{\mathbf{X}^{(k)}, \mathbf{X}_{k+1}}  \geq 1$, then $1 - \mathbf{R}_{\mathbf{X}^{(k)}, \mathbf{X}_{k+1}}^{t} \mathbf{R}^{(k),-1} \mathbf{R}_{\mathbf{X}^{(k)}, \mathbf{X}_{k+1}} \leq 0$, which is not possible since $\mathbf{R}{(k)}$ is a positive definite matrix.

Consequently, it has been verified that $1 - \mathbf{R}_{\mathbf{X}^{(k)}, \mathbf{X}_{k+1}}^{t} \mathbf{R}^{(k),-1} \mathbf{R}_{\mathbf{X}^{(k)}, \mathbf{X}_{k+1}}  < 1$ and, consequently, $| \mathbf{R}^{(k+1)} | < | \mathbf{R}^{(k)} |$.

That is to say, the determinant of $\mathbf{R}^{(k+1)}$ is less than that of $\mathbf{R}^{(k)}$, which is indicative of a greater degree of multicollinearity in the $\mathbf{X}^{(k+1)}$ matrix than in the $\mathbf{X}^{(k)}$ matrix (see, for example, García, Salmerón and García \cite{Garcia2019}).

\section{Correction coefficient tables}
    \label{appendix4}

This section includes the tables of the coefficients that weight the VIF (Table \ref{tab1a}), $t_{exp}$ (Table \ref{tab2a}) and $t_{n-k}(1-\alpha/2)$ (Table \ref{tab3a}).

\begin{table}[h]
    \centering
    {\small
    \begin{tabular}{cccccccccccccc}
        \hline
        $n$ vs $k$ & 3 & 4 & 5 & 6 & 7 & 8 & 9 & 10 & 11 & 12 & 13 & 14 & 15 \\
        \hline
        15 & 0.929 & 0.857 & 0.786 & 0.714 & 0.643 & 0.571 & 0.5 & 0.429 & 0.357 & 0.286 & 0.214 & 0.143 & 0.071 \\
        20 & 0.947 & 0.895 & 0.842 & 0.789 & 0.737 & 0.684 & 0.632 & 0.579 & 0.526 & 0.474 & 0.421 & 0.368 & 0.316 \\
        25 & 0.958 & 0.917 & 0.875 & 0.833 & 0.792 & 0.75 & 0.708 & 0.667 & 0.625 & 0.583 & 0.542 & 0.5 & 0.458 \\
        30 & 0.966 & 0.931 & 0.897 & 0.862 & 0.828 & 0.793 & 0.759 & 0.724 & 0.69 & 0.655 & 0.621 & 0.586 & 0.552 \\
        35 & 0.971 & 0.941 & 0.912 & 0.882 & 0.853 & 0.824 & 0.794 & 0.765 & 0.735 & 0.706 & 0.676 & 0.647 & 0.618 \\
        40 & 0.974 & 0.949 & 0.923 & 0.897 & 0.872 & 0.846 & 0.821 & 0.795 & 0.769 & 0.744 & 0.718 & 0.692 & 0.667 \\
        45 & 0.977 & 0.955 & 0.932 & 0.909 & 0.886 & 0.864 & 0.841 & 0.818 & 0.795 & 0.773 & 0.75 & 0.727 & 0.705 \\
        50 & 0.98 & 0.959 & 0.939 & 0.918 & 0.898 & 0.878 & 0.857 & 0.837 & 0.816 & 0.796 & 0.776 & 0.755 & 0.735 \\
        55 & 0.981 & 0.963 & 0.944 & 0.926 & 0.907 & 0.889 & 0.87 & 0.852 & 0.833 & 0.815 & 0.796 & 0.778 & 0.759 \\
        60 & 0.983 & 0.966 & 0.949 & 0.932 & 0.915 & 0.898 & 0.881 & 0.864 & 0.847 & 0.831 & 0.814 & 0.797 & 0.78 \\
        65 & 0.984 & 0.969 & 0.953 & 0.938 & 0.922 & 0.906 & 0.891 & 0.875 & 0.859 & 0.844 & 0.828 & 0.812 & 0.797 \\
        70 & 0.986 & 0.971 & 0.957 & 0.942 & 0.928 & 0.913 & 0.899 & 0.884 & 0.87 & 0.855 & 0.841 & 0.826 & 0.812 \\
        75 & 0.986 & 0.973 & 0.959 & 0.946 & 0.932 & 0.919 & 0.905 & 0.892 & 0.878 & 0.865 & 0.851 & 0.838 & 0.824 \\
        80 & 0.987 & 0.975 & 0.962 & 0.949 & 0.937 & 0.924 & 0.911 & 0.899 & 0.886 & 0.873 & 0.861 & 0.848 & 0.835 \\
        85 & 0.988 & 0.976 & 0.964 & 0.952 & 0.94 & 0.929 & 0.917 & 0.905 & 0.893 & 0.881 & 0.869 & 0.857 & 0.845 \\
        90 & 0.989 & 0.978 & 0.966 & 0.955 & 0.944 & 0.933 & 0.921 & 0.91 & 0.899 & 0.888 & 0.876 & 0.865 & 0.854 \\
        95 & 0.989 & 0.979 & 0.968 & 0.957 & 0.947 & 0.936 & 0.926 & 0.915 & 0.904 & 0.894 & 0.883 & 0.872 & 0.862 \\
        100 & 0.99 & 0.98 & 0.97 & 0.96 & 0.949 & 0.939 & 0.929 & 0.919 & 0.909 & 0.899 & 0.889 & 0.879 & 0.869 \\
        105 & 0.99 & 0.981 & 0.971 & 0.962 & 0.952 & 0.942 & 0.933 & 0.923 & 0.913 & 0.904 & 0.894 & 0.885 & 0.875 \\
        110 & 0.991 & 0.982 & 0.972 & 0.963 & 0.954 & 0.945 & 0.936 & 0.927 & 0.917 & 0.908 & 0.899 & 0.89 & 0.881 \\
        115 & 0.991 & 0.982 & 0.974 & 0.965 & 0.956 & 0.947 & 0.939 & 0.93 & 0.921 & 0.912 & 0.904 & 0.895 & 0.886 \\
        120 & 0.992 & 0.983 & 0.975 & 0.966 & 0.958 & 0.95 & 0.941 & 0.933 & 0.924 & 0.916 & 0.908 & 0.899 & 0.891 \\
        125 & 0.992 & 0.984 & 0.976 & 0.968 & 0.96 & 0.952 & 0.944 & 0.935 & 0.927 & 0.919 & 0.911 & 0.903 & 0.895 \\
        130 & 0.992 & 0.984 & 0.977 & 0.969 & 0.961 & 0.953 & 0.946 & 0.938 & 0.93 & 0.922 & 0.915 & 0.907 & 0.899 \\
        135 & 0.993 & 0.985 & 0.978 & 0.97 & 0.963 & 0.955 & 0.948 & 0.94 & 0.933 & 0.925 & 0.918 & 0.91 & 0.903 \\
        140 & 0.993 & 0.986 & 0.978 & 0.971 & 0.964 & 0.957 & 0.95 & 0.942 & 0.935 & 0.928 & 0.921 & 0.914 & 0.906 \\
        145 & 0.993 & 0.986 & 0.979 & 0.972 & 0.965 & 0.958 & 0.951 & 0.944 & 0.938 & 0.931 & 0.924 & 0.917 & 0.91 \\
        150 & 0.993 & 0.987 & 0.98 & 0.973 & 0.966 & 0.96 & 0.953 & 0.946 & 0.94 & 0.933 & 0.926 & 0.919 & 0.913 \\
        155 & 0.994 & 0.987 & 0.981 & 0.974 & 0.968 & 0.961 & 0.955 & 0.948 & 0.942 & 0.935 & 0.929 & 0.922 & 0.916 \\
        160 & 0.994 & 0.987 & 0.981 & 0.975 & 0.969 & 0.962 & 0.956 & 0.95 & 0.943 & 0.937 & 0.931 & 0.925 & 0.918 \\
        165 & 0.994 & 0.988 & 0.982 & 0.976 & 0.97 & 0.963 & 0.957 & 0.951 & 0.945 & 0.939 & 0.933 & 0.927 & 0.921 \\
        170 & 0.994 & 0.988 & 0.982 & 0.976 & 0.97 & 0.964 & 0.959 & 0.953 & 0.947 & 0.941 & 0.935 & 0.929 & 0.923 \\
        175 & 0.994 & 0.989 & 0.983 & 0.977 & 0.971 & 0.966 & 0.96 & 0.954 & 0.948 & 0.943 & 0.937 & 0.931 & 0.925 \\
        180 & 0.994 & 0.989 & 0.983 & 0.978 & 0.972 & 0.966 & 0.961 & 0.955 & 0.95 & 0.944 & 0.939 & 0.933 & 0.927 \\
        185 & 0.995 & 0.989 & 0.984 & 0.978 & 0.973 & 0.967 & 0.962 & 0.957 & 0.951 & 0.946 & 0.94 & 0.935 & 0.929 \\
        190 & 0.995 & 0.989 & 0.984 & 0.979 & 0.974 & 0.968 & 0.963 & 0.958 & 0.952 & 0.947 & 0.942 & 0.937 & 0.931 \\
        195 & 0.995 & 0.99 & 0.985 & 0.979 & 0.974 & 0.969 & 0.964 & 0.959 & 0.954 & 0.948 & 0.943 & 0.938 & 0.933 \\
        200 & 0.995 & 0.99 & 0.985 & 0.98 & 0.975 & 0.97 & 0.965 & 0.96 & 0.955 & 0.95 & 0.945 & 0.94 & 0.935 \\
        \hline
    \end{tabular}
    }
    \caption{Values of $a(n,k)$ for $n \in \{15, 20, 25,\dots, 195,200\}$ and $k \in \{3, 4, 5, \dots, 14, 15\}$.} \label{tab1a}
\end{table}

\begin{table}
    \centering
    {\small
    \begin{tabular}{cccccccccccccc}
        \hline
        $n$ vs $k$ & 3 & 4 & 5 & 6 & 7 & 8 & 9 & 10 & 11 & 12 & 13 & 14 & 15 \\
        \hline
        15 & 1.038 & 1.08 & 1.128 & 1.183 & 1.247 & 1.323 & 1.414 & 1.528 & 1.673 & 1.871 & 2.16 & 2.646 & 3.742 \\
        20 & 1.027 & 1.057 & 1.09 & 1.125 & 1.165 & 1.209 & 1.258 & 1.314 & 1.378 & 1.453 & 1.541 & 1.648 & 1.78 \\
        25 & 1.022 & 1.044 & 1.069 & 1.095 & 1.124 & 1.155 & 1.188 & 1.225 & 1.265 & 1.309 & 1.359 & 1.414 & 1.477 \\
        30 & 1.018 & 1.036 & 1.056 & 1.077 & 1.099 & 1.123 & 1.148 & 1.175 & 1.204 & 1.235 & 1.269 & 1.306 & 1.346 \\
        35 & 1.015 & 1.031 & 1.047 & 1.065 & 1.083 & 1.102 & 1.122 & 1.144 & 1.166 & 1.19 & 1.216 & 1.243 & 1.272 \\
        40 & 1.013 & 1.027 & 1.041 & 1.056 & 1.071 & 1.087 & 1.104 & 1.122 & 1.14 & 1.16 & 1.18 & 1.202 & 1.225 \\
        45 & 1.012 & 1.024 & 1.036 & 1.049 & 1.062 & 1.076 & 1.09 & 1.106 & 1.121 & 1.138 & 1.155 & 1.173 & 1.191 \\
        50 & 1.01 & 1.021 & 1.032 & 1.043 & 1.055 & 1.067 & 1.08 & 1.093 & 1.107 & 1.121 & 1.136 & 1.151 & 1.167 \\
        55 & 1.009 & 1.019 & 1.029 & 1.039 & 1.05 & 1.061 & 1.072 & 1.083 & 1.095 & 1.108 & 1.121 & 1.134 & 1.148 \\
        60 & 1.009 & 1.017 & 1.026 & 1.036 & 1.045 & 1.055 & 1.065 & 1.076 & 1.086 & 1.097 & 1.109 & 1.12 & 1.133 \\
        65 & 1.008 & 1.016 & 1.024 & 1.033 & 1.042 & 1.05 & 1.06 & 1.069 & 1.079 & 1.089 & 1.099 & 1.109 & 1.12 \\
        70 & 1.007 & 1.015 & 1.022 & 1.03 & 1.038 & 1.047 & 1.055 & 1.064 & 1.072 & 1.081 & 1.091 & 1.1 & 1.11 \\
        75 & 1.007 & 1.014 & 1.021 & 1.028 & 1.036 & 1.043 & 1.051 & 1.059 & 1.067 & 1.075 & 1.084 & 1.092 & 1.101 \\
        80 & 1.006 & 1.013 & 1.02 & 1.026 & 1.033 & 1.04 & 1.047 & 1.055 & 1.062 & 1.07 & 1.078 & 1.086 & 1.094 \\
        85 & 1.006 & 1.012 & 1.018 & 1.025 & 1.031 & 1.038 & 1.044 & 1.051 & 1.058 & 1.065 & 1.073 & 1.08 & 1.088 \\
        90 & 1.006 & 1.011 & 1.017 & 1.023 & 1.029 & 1.036 & 1.042 & 1.048 & 1.055 & 1.061 & 1.068 & 1.075 & 1.082 \\
        95 & 1.005 & 1.011 & 1.016 & 1.022 & 1.028 & 1.034 & 1.039 & 1.045 & 1.052 & 1.058 & 1.064 & 1.071 & 1.077 \\
        100 & 1.005 & 1.01 & 1.016 & 1.021 & 1.026 & 1.032 & 1.037 & 1.043 & 1.049 & 1.055 & 1.061 & 1.067 & 1.073 \\
        105 & 1.005 & 1.01 & 1.015 & 1.02 & 1.025 & 1.03 & 1.035 & 1.041 & 1.046 & 1.052 & 1.057 & 1.063 & 1.069 \\
        110 & 1.005 & 1.009 & 1.014 & 1.019 & 1.024 & 1.029 & 1.034 & 1.039 & 1.044 & 1.049 & 1.055 & 1.06 & 1.066 \\
        115 & 1.004 & 1.009 & 1.013 & 1.018 & 1.023 & 1.027 & 1.032 & 1.037 & 1.042 & 1.047 & 1.052 & 1.057 & 1.062 \\
        120 & 1.004 & 1.009 & 1.013 & 1.017 & 1.022 & 1.026 & 1.031 & 1.035 & 1.04 & 1.045 & 1.05 & 1.055 & 1.06 \\
        125 & 1.004 & 1.008 & 1.012 & 1.017 & 1.021 & 1.025 & 1.029 & 1.034 & 1.038 & 1.043 & 1.048 & 1.052 & 1.057 \\
        130 & 1.004 & 1.008 & 1.012 & 1.016 & 1.02 & 1.024 & 1.028 & 1.033 & 1.037 & 1.041 & 1.046 & 1.05 & 1.055 \\
        135 & 1.004 & 1.008 & 1.011 & 1.015 & 1.019 & 1.023 & 1.027 & 1.031 & 1.035 & 1.04 & 1.044 & 1.048 & 1.052 \\
        140 & 1.004 & 1.007 & 1.011 & 1.015 & 1.018 & 1.022 & 1.026 & 1.03 & 1.034 & 1.038 & 1.042 & 1.046 & 1.05 \\
        145 & 1.003 & 1.007 & 1.011 & 1.014 & 1.018 & 1.022 & 1.025 & 1.029 & 1.033 & 1.037 & 1.041 & 1.044 & 1.048 \\
        150 & 1.003 & 1.007 & 1.01 & 1.014 & 1.017 & 1.021 & 1.024 & 1.028 & 1.032 & 1.035 & 1.039 & 1.043 & 1.047 \\
        155 & 1.003 & 1.007 & 1.01 & 1.013 & 1.017 & 1.02 & 1.024 & 1.027 & 1.031 & 1.034 & 1.038 & 1.041 & 1.045 \\
        160 & 1.003 & 1.006 & 1.01 & 1.013 & 1.016 & 1.019 & 1.023 & 1.026 & 1.03 & 1.033 & 1.036 & 1.04 & 1.044 \\
        165 & 1.003 & 1.006 & 1.009 & 1.012 & 1.016 & 1.019 & 1.022 & 1.025 & 1.029 & 1.032 & 1.035 & 1.039 & 1.042 \\
        170 & 1.003 & 1.006 & 1.009 & 1.012 & 1.015 & 1.018 & 1.021 & 1.025 & 1.028 & 1.031 & 1.034 & 1.038 & 1.041 \\
        175 & 1.003 & 1.006 & 1.009 & 1.012 & 1.015 & 1.018 & 1.021 & 1.024 & 1.027 & 1.03 & 1.033 & 1.036 & 1.04 \\
        180 & 1.003 & 1.006 & 1.008 & 1.011 & 1.014 & 1.017 & 1.02 & 1.023 & 1.026 & 1.029 & 1.032 & 1.035 & 1.038 \\
        185 & 1.003 & 1.005 & 1.008 & 1.011 & 1.014 & 1.017 & 1.02 & 1.022 & 1.025 & 1.028 & 1.031 & 1.034 & 1.037 \\
        190 & 1.003 & 1.005 & 1.008 & 1.011 & 1.013 & 1.016 & 1.019 & 1.022 & 1.025 & 1.028 & 1.03 & 1.033 & 1.036 \\
        195 & 1.003 & 1.005 & 1.008 & 1.01 & 1.013 & 1.016 & 1.019 & 1.021 & 1.024 & 1.027 & 1.03 & 1.032 & 1.035 \\
        200 & 1.003 & 1.005 & 1.008 & 1.01 & 1.013 & 1.015 & 1.018 & 1.021 & 1.023 & 1.026 & 1.029 & 1.032 & 1.034 \\
        \hline
    \end{tabular}
    }
    \caption{Values of $b(n,k)$ for $n \in \{15, 20, 25,\dots, 195,200\}$ and $k \in \{3, 4, 5, \dots, 14, 15\}$.} \label{tab2a}
\end{table}

\begin{table}
    \centering
    {\small
    \begin{tabular}{cccccccccccccc}
        \hline
        $n$ vs $k$ & 3 & 4 & 5 & 6 & 7 & 8 & 9 & 10 & 11 & 12 & 13 & 14 & 15 \\
        \hline
        15 & 0.964 & 0.926 & 0.887 & 0.845 & 0.802 & 0.756 & 0.707 & 0.655 & 0.597 & 0.535 & 0.463 & 0.378 & 0.266 \\
        20 & 0.973 & 0.946 & 0.918 & 0.888 & 0.858 & 0.827 & 0.795 & 0.761 & 0.725 & 0.688 & 0.649 & 0.607 & 0.562 \\
        25 & 0.979 & 0.958 & 0.935 & 0.913 & 0.89 & 0.866 & 0.841 & 0.817 & 0.791 & 0.764 & 0.736 & 0.707 & 0.677 \\
        30 & 0.983 & 0.965 & 0.947 & 0.928 & 0.91 & 0.891 & 0.871 & 0.851 & 0.831 & 0.809 & 0.788 & 0.766 & 0.743 \\
        35 & 0.985 & 0.97 & 0.955 & 0.939 & 0.924 & 0.908 & 0.891 & 0.875 & 0.857 & 0.84 & 0.822 & 0.804 & 0.786 \\
        40 & 0.987 & 0.974 & 0.961 & 0.947 & 0.934 & 0.92 & 0.906 & 0.892 & 0.877 & 0.863 & 0.847 & 0.832 & 0.817 \\
        45 & 0.988 & 0.977 & 0.965 & 0.953 & 0.941 & 0.93 & 0.917 & 0.904 & 0.892 & 0.879 & 0.866 & 0.853 & 0.84 \\
        50 & 0.99 & 0.979 & 0.969 & 0.958 & 0.948 & 0.937 & 0.926 & 0.915 & 0.903 & 0.892 & 0.881 & 0.869 & 0.857 \\
        55 & 0.99 & 0.981 & 0.972 & 0.962 & 0.952 & 0.943 & 0.933 & 0.923 & 0.913 & 0.903 & 0.892 & 0.882 & 0.871 \\
        60 & 0.991 & 0.983 & 0.974 & 0.965 & 0.957 & 0.948 & 0.939 & 0.93 & 0.92 & 0.912 & 0.902 & 0.893 & 0.883 \\
        65 & 0.992 & 0.984 & 0.976 & 0.969 & 0.96 & 0.952 & 0.944 & 0.935 & 0.927 & 0.919 & 0.91 & 0.901 & 0.893 \\
        70 & 0.993 & 0.985 & 0.978 & 0.971 & 0.963 & 0.956 & 0.948 & 0.94 & 0.933 & 0.925 & 0.917 & 0.909 & 0.901 \\
        75 & 0.993 & 0.986 & 0.979 & 0.973 & 0.965 & 0.959 & 0.951 & 0.944 & 0.937 & 0.93 & 0.922 & 0.915 & 0.908 \\
        80 & 0.993 & 0.987 & 0.981 & 0.974 & 0.968 & 0.961 & 0.954 & 0.948 & 0.941 & 0.934 & 0.928 & 0.921 & 0.914 \\
        85 & 0.994 & 0.988 & 0.982 & 0.976 & 0.97 & 0.964 & 0.958 & 0.951 & 0.945 & 0.939 & 0.932 & 0.926 & 0.919 \\
        90 & 0.994 & 0.989 & 0.983 & 0.977 & 0.972 & 0.966 & 0.96 & 0.954 & 0.948 & 0.942 & 0.936 & 0.93 & 0.924 \\
        95 & 0.994 & 0.989 & 0.984 & 0.978 & 0.973 & 0.967 & 0.962 & 0.957 & 0.951 & 0.946 & 0.94 & 0.934 & 0.928 \\
        100 & 0.995 & 0.99 & 0.985 & 0.98 & 0.974 & 0.969 & 0.964 & 0.959 & 0.953 & 0.948 & 0.943 & 0.938 & 0.932 \\
        105 & 0.995 & 0.99 & 0.985 & 0.981 & 0.976 & 0.971 & 0.966 & 0.961 & 0.956 & 0.951 & 0.946 & 0.941 & 0.935 \\
        110 & 0.995 & 0.991 & 0.986 & 0.981 & 0.977 & 0.972 & 0.967 & 0.963 & 0.958 & 0.953 & 0.948 & 0.943 & 0.939 \\
        115 & 0.995 & 0.991 & 0.987 & 0.982 & 0.978 & 0.973 & 0.969 & 0.964 & 0.96 & 0.955 & 0.951 & 0.946 & 0.941 \\
        120 & 0.996 & 0.991 & 0.987 & 0.983 & 0.979 & 0.975 & 0.97 & 0.966 & 0.961 & 0.957 & 0.953 & 0.948 & 0.944 \\
        125 & 0.996 & 0.992 & 0.988 & 0.984 & 0.98 & 0.976 & 0.972 & 0.967 & 0.963 & 0.959 & 0.954 & 0.95 & 0.946 \\
        130 & 0.996 & 0.992 & 0.988 & 0.984 & 0.98 & 0.976 & 0.973 & 0.969 & 0.964 & 0.96 & 0.957 & 0.952 & 0.948 \\
        135 & 0.996 & 0.992 & 0.989 & 0.985 & 0.981 & 0.977 & 0.974 & 0.97 & 0.966 & 0.962 & 0.958 & 0.954 & 0.95 \\
        140 & 0.996 & 0.993 & 0.989 & 0.985 & 0.982 & 0.978 & 0.975 & 0.971 & 0.967 & 0.963 & 0.96 & 0.956 & 0.952 \\
        145 & 0.996 & 0.993 & 0.989 & 0.986 & 0.982 & 0.979 & 0.975 & 0.972 & 0.969 & 0.965 & 0.961 & 0.958 & 0.954 \\
        150 & 0.996 & 0.993 & 0.99 & 0.986 & 0.983 & 0.98 & 0.976 & 0.973 & 0.97 & 0.966 & 0.962 & 0.959 & 0.956 \\
        155 & 0.997 & 0.993 & 0.99 & 0.987 & 0.984 & 0.98 & 0.977 & 0.974 & 0.971 & 0.967 & 0.964 & 0.96 & 0.957 \\
        160 & 0.997 & 0.993 & 0.99 & 0.987 & 0.984 & 0.981 & 0.978 & 0.975 & 0.971 & 0.968 & 0.965 & 0.962 & 0.958 \\
        165 & 0.997 & 0.994 & 0.991 & 0.988 & 0.985 & 0.981 & 0.978 & 0.975 & 0.972 & 0.969 & 0.966 & 0.963 & 0.96 \\
        170 & 0.997 & 0.994 & 0.991 & 0.988 & 0.985 & 0.982 & 0.979 & 0.976 & 0.973 & 0.97 & 0.967 & 0.964 & 0.961 \\
        175 & 0.997 & 0.994 & 0.991 & 0.988 & 0.985 & 0.983 & 0.98 & 0.977 & 0.974 & 0.971 & 0.968 & 0.965 & 0.962 \\
        180 & 0.997 & 0.994 & 0.991 & 0.989 & 0.986 & 0.983 & 0.98 & 0.977 & 0.975 & 0.972 & 0.969 & 0.966 & 0.963 \\
        185 & 0.997 & 0.994 & 0.992 & 0.989 & 0.986 & 0.983 & 0.981 & 0.978 & 0.975 & 0.973 & 0.97 & 0.967 & 0.964 \\
        190 & 0.997 & 0.994 & 0.992 & 0.989 & 0.987 & 0.984 & 0.981 & 0.979 & 0.976 & 0.973 & 0.971 & 0.968 & 0.965 \\
        195 & 0.997 & 0.995 & 0.992 & 0.989 & 0.987 & 0.984 & 0.982 & 0.979 & 0.977 & 0.974 & 0.971 & 0.969 & 0.966 \\
        200 & 0.997 & 0.995 & 0.992 & 0.99 & 0.987 & 0.985 & 0.982 & 0.98 & 0.977 & 0.975 & 0.972 & 0.97 & 0.967 \\
        \hline
    \end{tabular}
    }
    \caption{Values of $\sqrt{a(n,k)}$ for $n \in \{15, 20, 25,\dots, 195,200\}$ and $k \in \{3, 4, 5, \dots, 14, 15\}$.} \label{tab3a}
\end{table}

\end{document}